\documentclass[%
reprint,
superscriptaddress,
nofootinbib,
 amsmath,amssymb,
 aps,
 prl,
nolongbibliography
]{revtex4-2}

\usepackage{graphicx}
\usepackage{dcolumn}
\usepackage{bm}
\usepackage[colorlinks=true
,urlcolor=blue
,anchorcolor=blue
,citecolor=blue
,filecolor=blue
,linkcolor=red
,menucolor=blue
linktocpage=true
pdfproducer=medialab
pdfa=true
]{hyperref}
\usepackage{nicefrac}
\usepackage{physics}
\usepackage{xspace}
\usepackage{xcolor}
\usepackage{newtxtext,newtxmath}
\usepackage[vlined]{algorithm2e}

\usepackage{orcidlink}

\definecolor{mygreen}{RGB}{20,138,6}
\definecolor{myblue}{RGB}{52,180,230}
\definecolor{myorange}{RGB}{230,120,10}

\newcommand{\vecb}[1]{\ensuremath{\boldsymbol{#1}}}
\providecommand{\dd}{\ensuremath{{\rm d}}}
\providecommand{\ee}{\ensuremath{{\rm e}}}
\providecommand{\ii}{\ensuremath{{\rm i}}}
\newcommand{\bnabla}{\ensuremath{\boldsymbol{\nabla}}}
\newcommand{\mPsi}{{\mathit \Psi}}
\newcommand{\mPi}{\mathit{\Pi}}

\usepackage[OT2,T1]{fontenc}
\DeclareSymbolFont{cyrletters}{OT2}{wncyr}{m}{n}
\DeclareMathSymbol{\Sha}{\mathalpha}{cyrletters}{"58}

\newcommand{\mysec}[1]{\vspace{0.1cm}\paragraph{\textbf{#1}.\textemdash\hspace{-0.25cm}}}

\providecommand{\ii}{\ensuremath{{\rm i}}}

\def\apjl{\ref@jnl{ApJ}}                

\begin{document}

\preprint{arXiv:2309.10865}

\title{Starting Cosmological Simulations from the Big Bang}

\author{Florian List\orcidlink{0000-0002-3741-179X}}
\email{florian.list@univie.ac.at (FL)}
\affiliation{Department of Astrophysics, University of Vienna, T\"{u}rkenschanzstra{\ss}e 17, 1180 Vienna, Austria}
\affiliation{Department of Mathematics, University of Vienna, Oskar-Morgenstern-Platz 1, 1090 Vienna, Austria}

\author{Oliver Hahn\orcidlink{0000-0001-9440-1152}}
\affiliation{Department of Astrophysics, University of Vienna, T\"{u}rkenschanzstra{\ss}e 17, 1180 Vienna, Austria}
\affiliation{Department of Mathematics, University of Vienna, Oskar-Morgenstern-Platz 1, 1090 Vienna, Austria}

\author{Cornelius Rampf\orcidlink{0000-0001-5947-9376}}
\affiliation{Department of Astrophysics, University of Vienna, T\"{u}rkenschanzstra{\ss}e 17, 1180 Vienna, Austria}
\affiliation{Department of Mathematics, University of Vienna, Oskar-Morgenstern-Platz 1, 1090 Vienna, Austria}

\date{\today}

\begin{abstract}
The cosmic large-scale structure (LSS) provides a unique testing ground for connecting fundamental physics to astronomical observations. 
Modeling the LSS requires numerical $N$-body simulations or perturbative techniques that both come with distinct shortcomings. Here we present the first unified numerical approach, enabled by new time integration and discreteness reduction schemes, and demonstrate its convergence at the field level. In particular, we show that our simulations (1) can be initialized directly at time zero, and (2) can be made to agree with high-order Lagrangian perturbation theory in the fluid limit. This enables fast, self-consistent, and UV-complete forward modeling of LSS observables.
\end{abstract}

\maketitle

\mysec{Introduction}
The gravitational evolution of collisionless matter is
governed by the cosmological Vlasov--Poisson system (VP; e.g.\ \cite{Peebles:1980, bernardeau2002large,Rampf:2021rqu,2022LRCA....8....1A}),
which describes how the phase-space distribution $f = f(t, \vecb{x}, \vecb{p})$ of a continuous medium evolves, 
\begin{equation}
    \frac{\dd f}{\dd t} = \frac{\partial f}{\partial t} + \frac{\vecb{p}}{a^2} \cdot \bnabla_{\vecb{x}} f - \bnabla_{\vecb{x}} \varphi \cdot \bnabla_{\vecb{p}} f = 0, \label{eq:vlasov-poisson}
\end{equation}
where the gravitational potential is subject to Poisson's equation $\nabla^2_{\vecb{x}} \varphi = 3/(2a) H_0^2 \Omega_m \delta$. Here, $(\vecb{x}, \vecb{p})$ is the canonical position-momentum pair, $a$ is the scale factor, $H_0$ is the Hubble constant, $\Omega_m$ is today's density parameter, and $\delta = \rho/\bar{\rho} - 1 = \int_{\mathbb{R}^3} f \, \dd^3 p - 1$ is the density contrast.

At sufficiently early times and assuming that matter is perfectly cold, the first two kinetic moments of Eq.~\eqref{eq:vlasov-poisson} form a closed set of fluid equations.
This resulting Euler--Poisson system is the starting point for perturbative approaches to structure formation, which form the basic \textit{theoretical} class of methods for studying the large-scale structure of the Universe: in Eulerian (standard) perturbation theory (e.g.\ \cite{bernardeau2002large}), the density contrast
$\delta$ is expanded in a Taylor series, and a hierarchy of recursion relations for $\delta$ is derived. However, as density fluctuations grow and $\delta \sim 1$, this technique breaks down.

An alternative approach is given by Lagrangian perturbation theory (LPT; e.g.\ \cite{Zeldovich:1970, Buchert:1993, Bouchet:1995, 1997GReGr..29..733E}), where instead a series ansatz is used for the displacement field $\vecb{\mPsi}(\vecb{q}) = \vecb{x}(\vecb{q}) - \vecb{q}$, i.e.\ the vector pointing from each Lagrangian position $\vecb{q}$ to the currently associated Eulerian position $\vecb{x}(\vecb{q})$ when moving along the fluid characteristics. 
All-order recursive solutions for $\vecb{\mPsi}$ are available \cite{Rampf:2012, Zheligovsky:2014, Matsubara:2015},
with the exact solution of the VP system arising in the limit of infinite order \cite{2018PhRvL.121x1302S}.
Although converging significantly faster than Eulerian perturbation theory, LPT eventually also breaks down, namely at the first shell-crossing, i.e.\ when particle trajectories cross for the first time. Then, the (Eulerian) velocity field becomes multi-valued, and the fluid description ceases to be valid as the Vlasov hierarchy can no longer be truncated at first order. Analytical post-shell-crossing approaches exist (e.g. \cite{Colombi2015, Taruya2017, Rampf:2021, 2022A&A...664A...3S}); however, they do not (yet) extend into the strongly non-linear regime and are therefore not mature enough to be useful in practice.
An alternative to this is e.g.\ `effective field theory of large-scale structure' \cite{baumann2012cosmological, carrasco2012effective, cabass2023snowmass}, which however relies on matching free parameters to a UV-complete approach, typically provided by simulations.

Hence, resolving the non-linear late-time dynamics in a UV-complete manner requires \textit{numerical} methods, with the most prominent technique given by $N$-body simulations. Here, the continuous phase-space distribution $f$ is represented by a set of $N$ discrete tracer particles with canonical positions and momenta $(\vecb{X}_i, \vecb{P}_i)$, for $i = 1, \ldots, N$. Requiring $\dd f(t, \vecb{X}_i, \vecb{P}_i) / \dd t = 0$ leads to the Hamiltonian equations of motion $\dot{\vecb{X}}_i = \vecb{P}_i / a^2$ and $\dot{\vecb{P}}_i = - \bnabla_{\vecb{x}} \varphi |_{\vecb{X}_i}$. Note that if one had access to the exact (continuous) potential $\varphi$, the particles would move \textit{exactly} according to the characteristics of the underlying continuous system. However, since the true density contrast $\delta$ in the Poisson equation can only be approximated based on the positions of the $N$ particles, an estimate $\delta_N \approx \delta$ sources the Poisson equation, resulting in an approximate potential $\varphi_N \approx \varphi$. This is the crucial approximation made by $N$-body simulations and, as we will see later, carefully designed techniques to improve the match $\varphi_N \to \varphi$ are therefore key in suppressing discreteness effects at early times and accessing the fluid limit with $N$-body simulations.

Although perturbation theory and $N$-body simulations are the theoretical and numerical pillars of modeling cosmological structure formation,
there are only very few studies on their agreement in the fluid limit at early times when perturbation theory is still valid. Comparison studies in this regime are hampered by the fact that spurious discreteness effects become significant at early times as the $N$-body system quickly deviates from the underlying continuous dynamics \cite{Joyce:2005, Marcos:2006}. Techniques for correcting at least the linear discreteness error of the particle lattice exist \cite{Garrison:2016}, but are not widely employed. Despite discreteness errors, $N$-body simulations have traditionally been initialized using first-order LPT (the Zel'dovich approximation \cite{Zeldovich:1970}) or, more recently, second-order LPT (2LPT \cite{1992ApJ...394L...5B,Buchert:1993}) at early times (redshift $z = a^{-1} - 1 \gtrsim 100$), to avoid truncation errors arising from the residual between LPT and the true solution, which ultimately bias the statistics of the simulated fields.
In Ref.~\cite{Michaux:2021}, it was recently shown that
a more favorable trade-off between numerical discreteness errors and LPT truncation errors is achieved by initializing cosmological simulations at rather \textit{late} times (e.g.\ $z \approx 15 - 40$), by employing higher-order LPT, namely 3LPT \cite{1994MNRAS.267..811B,1995MNRAS.276..115C}.

In this \textit{letter}, we bridge the gap between the analytical and numerical descriptions of cosmic structure formation in the fluid limit at early times. Specifically, we show for the first time that by applying an array of discreteness reduction techniques, together with a time integrator that has the correct asymptotic behavior for $a \to 0$, one obtains excellent agreement between $N$-body dynamics and perturbation theory. The choice of appropriate initial conditions (ICs) and time variable allows us to initialize $N$-body simulations at $a = 0$, enabling a clean comparison between $N$-body and LPT dynamics. Remarkably, a \textit{single} $N$-body drift-kick-drift (DKD) step from $a = 0$ to a `typical' 3LPT initialization time for cosmological simulations yields a displacement field at $\approx$~3LPT accuracy. This effectively renders moot the LPT-based initialization of cosmological $N$-body simulations and demonstrates that starting them directly at $a = 0$ is a promising alternative.

The structure of this \textit{letter} is as follows. First, we briefly review the time integrator \textsc{PowerFrog}, which we recently introduced in Ref.~\cite{list2023perturbation}. This integrator is asymptotically consistent with 2LPT for $a \to 0$ and a crucial ingredient for achieving agreement between LPT and the $N$-body dynamics. Next, we describe the discreteness suppression techniques that enable us to achieve extremely low-noise results in the fluid limit at early times. Then, we present and discuss our results for a single $N$-body simulation step from $z = \infty$ to $z = 18$ (shortly before the time of the first shell-crossing).
Finally, we comment on the present-day (i.e.\ $z = 0$) statistics of $N$-body simulations initialized either directly at $z = \infty$ or with LPT. We find that while the power spectra match to within $1\%$ even without applying any discreteness suppression techniques, these techniques are necessary in order to obtain the correct cross-power spectrum with $z = \infty$-initialized simulations.

\mysec{$\vecb{\mPi}$-integrators}
The leapfrog integrator is ubiquitous in cosmological simulations thanks to its simplicity, symplecticity, and suitability for individual time steps for different particles (e.g.\ \cite{HockneyEastwood}). While it converges at second order towards the correct solution as the time step decreases, it does not exploit the fact that before shell-crossing the displacement field $\vecb{\mPsi}$ can be expressed analytically in the form of a series in the linear growth-time $D$ of the $\Lambda$CDM concordance model, namely the LPT series $\vecb{\mPsi}(\vecb{q}, D) = \sum_{n=1}^\infty \vecb{\psi}^{(n)}(\vecb{q}) \, D^n$.%
\footnote{We only consider growing-mode solutions and neglect higher-order LPT corrections stemming from the cosmological constant $\Lambda$; see Refs.\ \cite{Rampf:2022,2022PhRvD.106l3504F}.}

In Ref.~\cite{list2023perturbation}, we introduced a class of integrators, which we named $\mPi$-integrators in view of the momentum variable $\vecb{\mPi} = \dd \vecb{X} / \dd D$ with respect to which they are formulated.
Expressing the integrator in terms of momentum $\vecb{\mPi}$ w.r.t.\ growth-factor time enables the construction of second-order accurate integration schemes which, when performing only few time steps, mimic LPT dynamics.

The only previously existing representative of this class is the popular \textsc{FastPM} scheme by Ref.~\cite{Feng:2016}, which was constructed to match the dynamics of the Zel'dovich approximation on large scales. One of our new integrators, which we named \textsc{PowerFrog}, further matches the 2LPT asymptote at early times $a \to 0$, which turns out to be essential for initializing simulations at $a = 0$, as we will see later.

As usual, we choose the ICs to be $\delta(D = 0) = 0$ and $\vecb{\mPi}(D = 0) = - \bnabla_{\vecb{q}} \phi_{\text{ini}}$, which implicitly selects the growing-mode solution and ensures that the initial momentum is curl-free \cite{Brenier2003, 2002Natur.417..260F, Rampf2019QuasiSpherical}; see e.g.\ Ref.\,\cite{Michaux:2021} for details on how $\phi_{\text{ini}}$ can be obtained from standard Boltzmann code employing a standard backscaling approach.
Notice that the canonical variables $(\vecb{X}, \vecb{P})$ are incompatible with these ICs: due to Liouville's theorem for Hamiltonian mechanics, the contraction of the positions to a single point in the limit $a \to 0$ leads to the divergence of the momenta. This is not so, however, for the coordinates $(\vecb{X}, \vecb{\mPi})$, which are employed by $\mPi$-integrators. In fact, the transformation from $(\vecb{X}, \vecb{P}) \mapsto (\vecb{X}, \vecb{\mPi})$ is non-canonical (but rather `contact', see \cite{Arnold:1989, Bravetti2017a}), for which reason these new variables are not subject to Liouville's theorem, and it is easy to see that the contact Hamiltonian for $(\vecb{X}, \vecb{\mPi})$ remains bounded for $a \to 0$ (subject to suitable ICs \cite{Rampf2021TwoFluids}). 

Equipped with an integrator that works in terms of these variables and, by construction, is consistent with the 2LPT trajectory at early times, we will demonstrate that it is possible to start cosmological simulations at \mbox{$a = 0$,} with the particles placed on an unperturbed homogeneous grid (which approximates $\delta(D = 0) = 0$), and the growth-factor `Zel'dovich' momentum initialized as $\vecb{\mPi}_i = -\bnabla_{\vecb{q}} \phi_{\text{ini}}|_{\vecb{X}_i}$. 

We emphasize that---in contrast to LPT---the time integration of cosmological $N$-body systems using $\mPi$-integrators is UV complete in that the $N$-body dynamics should converge towards the VP solution in the limit of infinitely many particles and time steps, even in the highly non-linear multistreaming regime (albeit the mathematical proof thereof is still missing; but see e.g.\ \cite{2021A&A...647A..66C,2023arXiv230706146F}).

\mysec{Towards the fluid limit: suppressing particle noise with sheet-based interpolation}
\label{sec:resampling}

In this \emph{letter}, we focus on the particle-mesh (PM) method for the force computation, but we also briefly consider Tree-PM \cite{bagla2002treepm, Bode2003} and the non-uniform Fast Fourier Transform (FFT). 
To control particle discreteness effects at the required level, we apply \textit{four} important steps:
\begin{enumerate} \itemsep0em 
    \item Increasing the number of gravity source particles (`resampling') by \textit{sheet interpolation} during the force calculation (extending the quadratic interpolation of Ref.~\cite{Hahn:2016} to Fourier interpolation, see also Ref.~\cite{stucker2018median})
    \item Using \textit{higher-order mass-assignment schemes} to represent particle positions more accurately \cite{Chaniotis:2004} and deconvolving the density field on the grid with the mass assignment kernel
    \item Using \textit{grid interlacing} to suppress low-order aliases \cite{HockneyEastwood}
    \item Using the \textit{exact gradient kernel} $\text{i}\vecb{k}$ for the force calculation, rather than a finite difference gradient kernel.
\end{enumerate}
The sheet interpolation has by far the largest effect in terms of suppressing discreteness. It harnesses the fact that for cold ICs, the phase-space density $f(t, \vecb{x}, \vecb{p})$ in the VP equation~\eqref{eq:vlasov-poisson} occupies a 3-dimensional manifold in the 6-dimensional phase space at all times, the Lagrangian submanifold.
Hence, to increase the spatial resolution of the gravitational potential, we can `spawn' new $N$-body gravity source particles in Lagrangian space on a finer grid, determine their displacement by Fourier interpolation, and compute the resulting force on the $N$ particles using this refined potential field. 

\begin{figure}
    \centering
    \resizebox{\columnwidth}{!}{
    \includegraphics{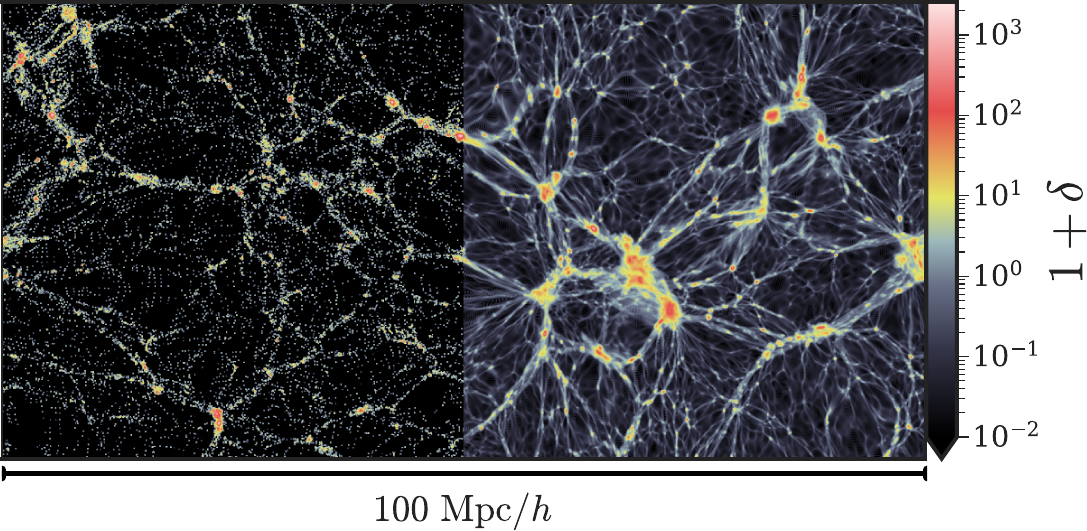}
    }
    \caption{Slice through the overdensity field at $z = 0$ of a standard $N$-body simulation (\textit{left half}) and a simulation with discreteness reduction techniques applied (in each time step, as well as for the computation of the plotted density slice, \textit{right half}). The particle and grid resolutions for both cases are $N = 512^3$ and $M = 1024^3$.}
    \label{fig:resampling-plot}
\end{figure}

For illustration, Fig.~\ref{fig:resampling-plot} shows a plot of the $z = 0$ density field with $N = 512^3$ particles, evaluated on a grid with $M = 1024^3$ cells, from a standard $N$-body simulation (\textit{left half}) and a simulation with $5^3$-fold resampling of each particle for the density computation and the other discreteness reduction techniques applied in each simulation step (\textit{right half}). The density field in the standard simulation is poorly sampled, particularly in underdense regions, with many cells containing no particles and hence $\delta_N = -1$. The resampling evidently suppresses discreteness and leads to a much more continuous density field.\footnote{Due to increasing complexity, the sheet-based interpolation \cite{Hahn:2016, stucker2020simulating} should not be applied in halo regions without using any refinements. 
Figure~\ref{fig:resampling-plot} is intended for illustrative purposes; here,
we employ resampling only in the fluid regime at early times when it is well suited to suppress discreteness.}
For a detailed explanation of each of these techniques, we refer to the Supplementary Material.

\mysec{Initializing simulations without LPT}

\begin{figure*}
    \centering
    \resizebox{0.9\textwidth}{!}{
    \includegraphics{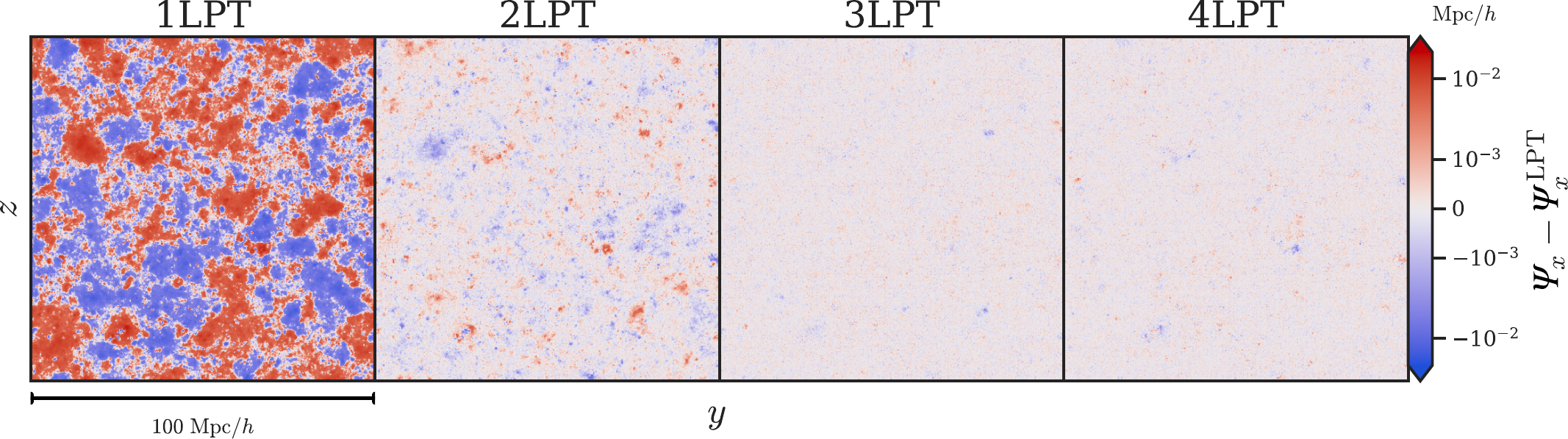}
    }
    \caption{Residuals of the $x$-displacement $\mPsi_x$ between our results with a \textit{single} \textsc{PowerFrog} $N$-body step from redshift $z = \infty$ to $z = 18$ and the corresponding LPT fields at $z = 18$ for different LPT orders. Shown is a slice in the Lagrangian $y$-$z$ coordinate plane.
    }
    \label{fig:displacement_plot}
\end{figure*}

We will now perform a single \textsc{PowerFrog} DKD time step starting from $z = \infty$ (i.e.\ $a = 0$) to a redshift where one would typically initialize a cosmological simulation with 3LPT, namely $z = 18$.  We checked that the Jacobi determinant $\det\left(\dd \vecb{x}^{\text{1LPT}} / \dd \vecb{q}\right) > 0$ for all particles at that time, and the standard deviation of the density field $\sigma(\delta^{\text{1LPT}}) = 0.30$. Hence, the entire simulation box is still in the single-stream regime, 
for which there is strong evidence that LPT converges \cite{2018PhRvL.121x1302S, Rampf:2021}.

We consider the evolution of $N = 512^3$ particles in a periodic simulation box of edge length $L = 100 \ \text{Mpc} / h$ subject to a flat $\Lambda$CDM cosmology with $\Omega_m = 0.3$, $H_0 = 67.11 \ \text{km} / \text{s} / \text{Mpc}$, $n_s = 0.9624$, $\sigma_8 = 0.8$. We perform our computations on a single GPU, computing the forces with the PM method at grid resolution $M = 1024^3$.

Figure~\ref{fig:displacement_plot} depicts the residual between the 1-step $N$-body result and different LPT orders at $z = 18$; specifically, we show a slice of the displacement component $\mPsi_x$. In view of \textsc{PowerFrog} being designed to only match the 2LPT asymptote for $a \to 0$, it might surprise that the 1-step $N$-body displacement lies even closer to 3LPT and 4LPT than to the 2LPT result. Intuitively, this can be understood by noting that the LPT terms are computed at the Lagrangian particle positions, that is, by `pulling back the evolution of each particle to its initial location, while the kick in the $N$-body step updates the velocities at growth-factor time $\Delta D / 2$ directly based on the potential that solves the Poisson equation at that time, which excites higher-order LPT terms; we leave a detailed investigation on this for future work.

We remark that also the velocity field is in good agreement with its LPT counterpart (see the Supplementary Material).
The excellent match between the positions and momenta of a single \textsc{PowerFrog} step and high-order LPT makes the initialization of cosmological simulations directly at the origin of time at $a = 0$ with \textsc{PowerFrog} (or another integrator that follows the $\geq 2$LPT asymptotic behavior for $a \to 0$) an attractive alternative to the traditional LPT-based ICs.

\begin{figure}[b]
    \centering
    \resizebox{1\columnwidth}{!}{
    \includegraphics{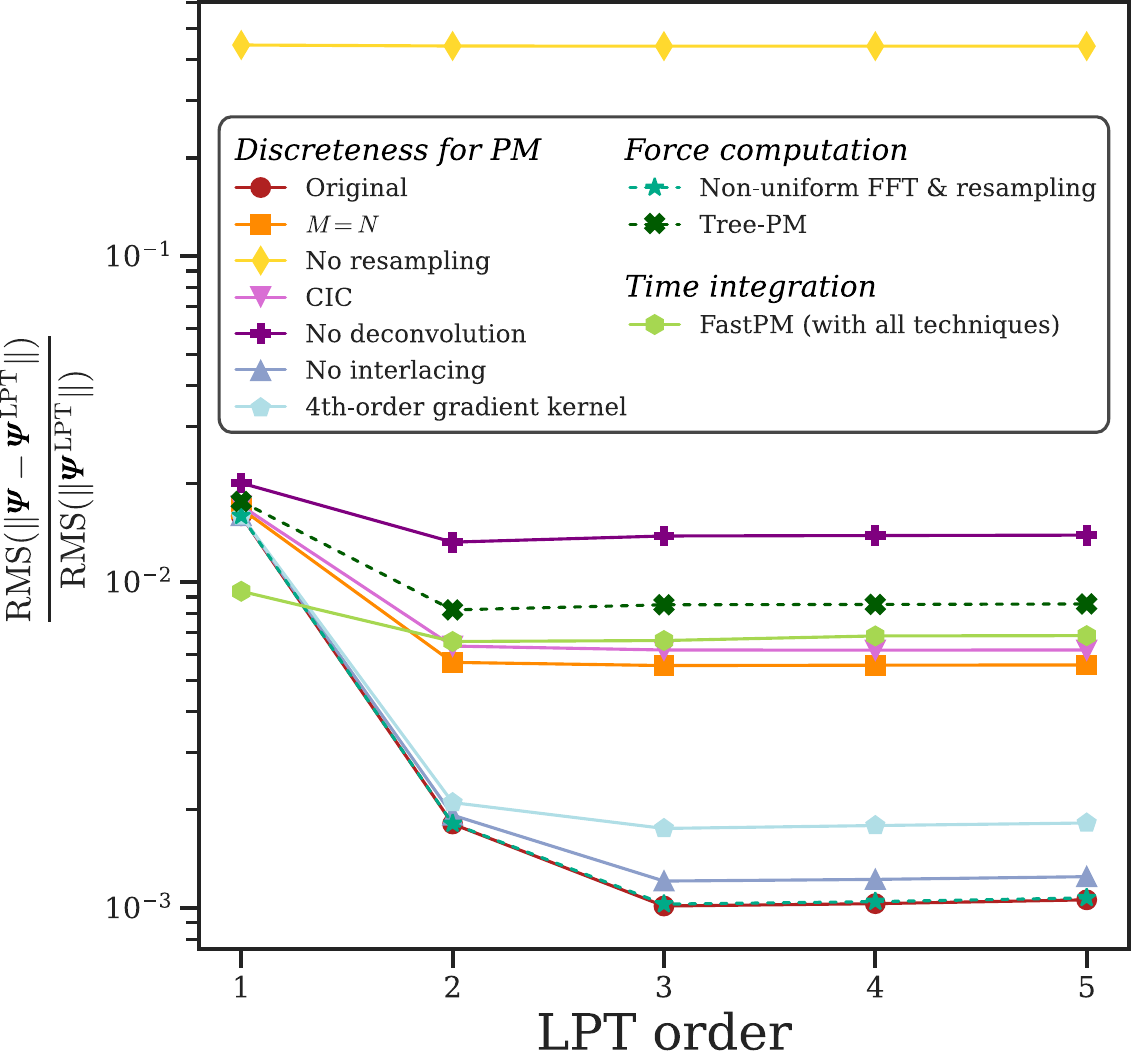}
    }
    \caption{Relative RMS displacement error between the 1-step $N$-body simulation and different LPT orders at $z = 18$ when using the \textsc{PowerFrog} integrator and applying all discreteness reduction techniques (`Original'), when omitting one technique at a time, and when performing a \textsc{FastPM} DKD step instead of a \textsc{PowerFrog} step (applying all discreteness reduction techniques). Evidently, a carefully designed time integrator, resampling, deconvolution, higher-order mass assignment, and a fine PM grid (e.g.\ $M = 2^3 N$) are all key ingredients to access the $3$LPT regime. We also show the residuals when using the non-uniform FFT instead of local mass assignment (with resampling), and with Tree-PM force computation (without resampling).}
    \label{fig:psi_rms_plot}
\end{figure}

We now study how the different discreteness reduction techniques affect the numerical solution of the $N$-body simulation.
Figure~\ref{fig:psi_rms_plot} shows the relative root-mean-square (RMS) error of the displacements between a single $N$-body step from $z = \infty$ to $z = 18$ using all discreteness suppression methods discussed above, together with the results when omitting one of these techniques at a time. Clearly, the sheet-based resampling of the density is crucial for achieving convergence between $N$-body and LPT: without it, the residual towards LPT is entirely dominated by errors due to the particle-based approximation of the continuous density at a level of $\sim 50\%$, and no differences between the different LPT orders are visible. The second-most important technique is the deconvolution of the density with the mass assignment kernel, whose absence results in significant high-frequency noise that conceals the $3-4$LPT contributions in the residual. The residual also increases substantially when reducing the number of PM grid cells from $M = 2^3 N$ to $M = N$ or when using cloud-in-cell (CIC) instead of piecewise-cubic spline (PCS) mass assignment. The impact of dealiasing the density by means of interlaced grids, and of using the exact Fourier gradient kernel $\ii \vecb{k}$ instead of a 4$^\text{th}$-order finite-difference gradient kernel is much more modest; however, leaving out any of these methods imprints a characteristic grainy structure in the $3-4$LPT residuals. With all techniques active, the 3LPT vs. 1-step \textsc{PowerFrog} residual is only 0.1\%.
For completeness, we also show the residuals when replacing the local PM mass assignment by Ref.~\cite{jaxfinufft}'s implementation of the non-uniform FFT (\cite{barnett2019parallel, barnett2021aliasing, shih2021cufinufft}) together with resampling, which yields the same residuals as our PM baseline and could be an exciting avenue for future exploration. Using Tree-PM for the force computation (without resampling) rather than PM also somewhat reduces discreteness, but much less than in our discreteness-suppressed PM baseline.

Finally, the green hexagons show the residual when performing a single step with the (DKD variant of the) \textsc{FastPM} stepper instead of \textsc{PowerFrog} which, recall, is consistent with the Zel'dovich approximation, but whose asymptotic behavior for $a \to 0$ differs from 2LPT. Clearly, there is a significant 2LPT contribution in the residual, which prevails in the residuals w.r.t.\ higher LPT orders. A plot of the residual fields and their power spectra can be found in the Supplementary Material. \textsc{FastPM} is therefore not suitable as a 1-step initializer.

In principle, it should be possible to construct integrators that match even higher LPT orders with a single time step by composing each step out of more than three drift/kick components, but the gain from going beyond 3LPT can be expected to be relatively small in practical applications. After the first time step, the assumption that the time step starts in the asymptotic regime at $a \approx 0$ is no longer exactly valid, and second-order-in-$a$ residuals arise with \textsc{PowerFrog}; we will explore potential improvements in this regard in future work. 

\mysec{Analysis at $\boldsymbol{z = 0}$}
Finally, let us comment on the results one obtains when using the positions and momenta computed with a single \textsc{PowerFrog} step as the ICs for a (standard) cosmological simulation down to $z = 0$. We take the same cosmology as in the previous section, $N = 512^3$ particles, and initialize the simulations either with 1, 2, or 3LPT, or with a single $N$-body time step (that starts from $z = \infty$) at $z = 36$; 
then, we run a cosmological simulation with the industry standard code \textsc{Gadget-4} \cite{Gadget4}. For the single $N$-body step, we consider \textsc{PowerFrog} (i) with discreteness suppression and grid sizes $M = 512^3$ and $1024^3$, (ii) without any discreteness suppression and $M = 512^3$, (iii) a \textsc{FastPM} DKD step with discreteness suppression, (iv) non-uniform FFT forces with resampling, and (v) Tree-PM forces (without resampling).

\begin{figure}
    \centering
    \resizebox{1\columnwidth}{!}{
    \includegraphics{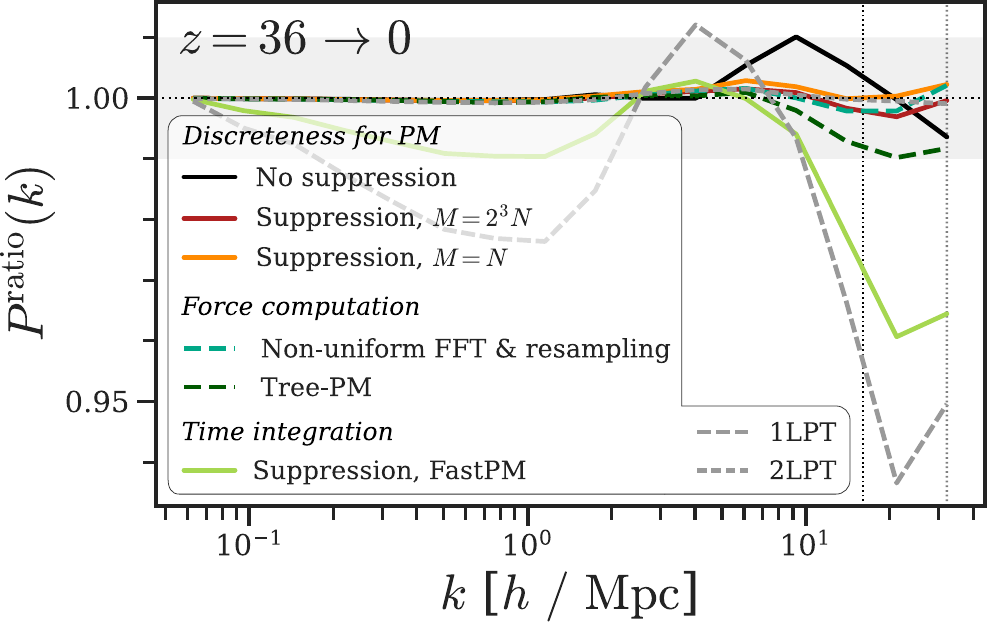}
    }
    \caption{
    Density power spectrum ratio at $z = 0$ between cosmological simulations with 1LPT, 2LPT, or 1-step $N$-body ICs and 3LPT ICs. Even without any discreteness suppression, starting cosmological simulations with a single \textsc{PowerFrog} step from $z = \infty$ to $36$ and then continuing in standard $N$-body fashion leads to power spectrum errors $< 1\%$ towards 3LPT on all scales. The dotted vertical lines show the Nyquist modes for $N = 512^3$ and $M = 1024^3$, respectively.}
    \label{fig:pk_plot_z_0}
\end{figure}

Figure~\ref{fig:pk_plot_z_0} shows the power spectrum ratio at $z = 0$ w.r.t.\ the $3$LPT ICs, which we take as our reference. Interestingly, even without any discreteness suppression, the residual between the power spectra with \textsc{PowerFrog} and 3LPT ICs is $\leq 1\%$ on all scales. Also, for the equilateral bispectrum, we find excellent agreement; however, the cross-spectrum drops significantly when omitting the discreteness reduction (e.g., from $99\%$ to $90\%$ at $k = 21 \ h / \text{Mpc}$, see the Supplementary Material). This implies that in principle, standard $N$-body simulations can be started with a \textsc{PowerFrog}-like stepper without any discreteness suppression, and the resulting $z = 0$ density field will be correct in terms of power spectrum and bispectrum, but its phases will be somewhat corrupted due to the discreteness.

\mysec{Discussion}
In this \textit{letter}, we have provided the first demonstration of the field-level agreement between high-order LPT and cosmological $N$-body simulations in the single-stream regime. Choosing kinematic variables in which the solution remains regular for $a \to 0$ allowed us to initialize simulations at the origin of time, making the customary LPT-based computation of the initial conditions at some scale factor $a \!>\!0\,$ obsolete---provided discreteness artifacts are sufficiently suppressed. 
Remarkably, the use of an LPT-informed time integrator implies that a \textit{single} $N$-body step starting from $a = 0$ yields more accurate results than 2LPT, which is the established technique for initializing cosmological simulations. 

From a practical point of view, this opens up a wide range of applications: the computational cost of the discreteness-suppressed step we applied to obtain the close match with LPT at early times shown in Fig.~\ref{fig:displacement_plot} is similar to that of 3LPT, but already applying very few or even none of these techniques might give sufficiently accurate results in fast simulations and for analyses focused on late times (see the highly accurate power spectrum for `No suppression' in Fig.~\ref{fig:pk_plot_z_0}). Also, an $N$-body initialization step from $a = 0$ might be superior in terms of memory requirements---no matter how fine the grid in Lagrangian space used for the resampling---as no large arrays need to be stored for each LPT order.
Another interesting scope of application is given by zoom simulations, where the intricacies in the (usually FFT-based, but cf.~\cite{Garrison:2016} for 2LPT computed in configuration space) LPT computation arising from different resolutions can be circumvented.

In the era of precision cosmology, it is crucial to thoroughly test the agreement of complementary approaches to structure formation such as perturbative techniques and numerical methods and to clearly identify their range of validity.
Our findings in this \textit{letter} lay the groundwork for further comparison studies at the intersection between analytical and numerical methods.

\begin{acknowledgments}
{\it We thank Raul Angulo and Jens St\"{u}cker for insightful discussions.~OH thanks Tom Abel for many past discussions on discreteness and the sheet. A software package implementing the discussed algorithms will be released in the near future.
}
\end{acknowledgments}


\def\mnras{Mon. Not. R. Astron. Soc.}\def\mnrasl{Mon. Not. R. Astron. Soc.
  Letters}\def\aap{Astron. Astrophys.}\def\prd{Phys. Rev. D}\def\prl{Phys. Rev.
  Lett.}\def\jcap{ J. Cosmol. Astropart. Phys.}\def\nat{Nature}\def\jcp{J.
  Comput. Phys.}\def\apj{Astrophys. J.}\def\apjs{Astrophys. J. Supp.
  Ser.}\def\preprint{Preprint }

\renewcommand{\thefigure}{S\arabic{figure}}
\setcounter{figure}{0}
\renewcommand{\theequation}{S\arabic{equation}}
\setcounter{equation}{0}
\newpage
\onecolumngrid

\section{Asymptotics at the Big Bang Singularity, Perturbation Theory, and \texorpdfstring{$\boldsymbol{N}$-body}{N-body} Time Integration}

\noindent Consider particle trajectories $a\mapsto \vecb{X}(\vecb{q},a)\in\mathbb{T}^3:=\mathbb{R}^3/\mathbb{Z}^3$ parameterized by the scale factor $a$ of the Universe and indexed by~$\vecb{q}\in\mathbb{T}^3$ (without loss of generality, this assumes units in which the box size is unity). For simplicity, let us assume Einstein-de~Sitter asymptotics for $a\to0$ in this section.

Following Ref.~\cite{Brenier:2003}, the equations of motion can be written as
\begin{align}
    \frac{2a}{3} \frac{\dd^2\vecb{X}}{\dd a^2} + \frac{\dd \vecb{X}}{\dd a} + \bnabla_{\vecb{x}} \phi(\vecb{X}, a) &= 0 & 1 + \frac{a}{H_0^2} \nabla^2_{\vecb{x}} \phi &= 1 + \delta = \int_{\mathbb{T}^3}\dd^3q\;\delta_{\rm D}(\vecb{x}-\vecb{X}(\vecb{q},a)),\label{eq:EOM}
\end{align}
where we defined the rescaled gravitational potential $\phi := 2 \varphi / 3$ for convenience.

\subsection{Asymptotics for \texorpdfstring{$\boldsymbol{a\to0}$}{a to 0} and perturbation theory}

\noindent In the limit $a\to0$, the equations of motion \eqref{eq:EOM} impose the following asymptotic constraints (referred to as `slaving' by \cite{Brenier:2003})
\begin{align}
    \frac{\dd \vecb{X}}{\dd a} &\asymp -\bnabla_{\vecb{q}} \phi_{\text{ini}} & \delta &\asymp 0 \quad\Leftrightarrow\quad \vecb{X}(\vecb{q},a) \asymp \vecb{q}
\end{align}
as initial conditions, where $\phi_{\text{ini}} := \phi(\vecb{q}, 0)$. These initial constraints therefore require the density of the universe to become asymptotically uniform, and particle velocities to be of purely potential nature with a single scalar degree of freedom $\phi_{\text{ini}}$ remaining, which describes the entire initial condition of the universe. The $a\to0$ asymptotics of the equations of motion therefore enforce the Zel'dovich approximation 
\begin{align}
    \vecb{X}(\vecb{q},a) = \vecb{q} - a \bnabla_{\vecb{q}} \phi_{\text{ini}} + \mathcal{O}(a^2)
\end{align}
at leading order for $0\leq a\ll 1$.

At higher orders, it is customary in so-called Lagrangian perturbation theory (LPT) to expand the displacement field in terms of a Taylor series in a suitable time variable with space-dependent Taylor coefficients, i.e.
\begin{align}
    \vecb{\mPsi}(\vecb{q},a) := \vecb{X}(\vecb{q},a) - \vecb{q} = \sum_{n=1}^\infty \vecb{\psi}^{(n)}(\vecb{q}) \; a^n,
\end{align}
for which all-order recursive relations are known \cite{Rampf:2012, Zheligovsky:2014} along with corrections for realistic $\Lambda$CDM cosmologies \cite{Matsubara:2015,Rampf:2022,Fasiello:2022}. 

LPT is by construction limited in its applicability to the regime $a<a_\ast$, where $a_\ast$ is the moment when particle trajectories begin to overlap (i.e.\ the flow field becomes multi-kinetic). This moment is associated with sign-flips of the Jacobian of the Lagrangian map, i.e.
\begin{align}
  \det \left( \frac{\dd\vecb{X}(\vecb{q},a)}{\dd\vecb{q}} \right) > 0 \quad \text{for all} \; \vecb{q}\in\mathbb{T}^3\;\text{and}\;a<a_\ast.
\end{align}
It is well known that LPT converges fairly quickly even at times close to $a_\ast$,
which justifies the use of low-order LPT truncations \cite{Michaux:2021,Rampf:2021}, except near regions that are very close to being spherically symmetric \cite{Nadkarni-Ghosh:2011,Rampf:2023,2023arXiv230312832R}. For this reason, it has been the method of choice to provide accurate initial conditions for cosmological simulations \cite{Crocce2006,Michaux:2021}.

A distinct disadvantage of (standard) LPT is that all evaluations of gravitational interactions must be carried out in Lagrangian space, requiring push~forward of particles followed by pull~back perturbative expansions. The distinct {\it advantage} of LPT over $N$-body simulations is, however, that the calculation is carried out in the \textit{fluid limit}, i.e.\ the fluid elements are not discretized in principle. In practice, when used in the context of cosmological simulations, a discrete set of modes, truncated in the UV, is of course employed. Still, calculations \cite{Joyce:2005,Marcos:2006} and numerical experiments \cite{Garrison:2016,Michaux:2021} have shown that $N$-body simulations do not agree with the fluid-limit evolution. To remedy such effects, Ref.~\cite{Garrison:2016} has proposed to correct the particle motion at linear order for the error, while Ref.~\cite{Michaux:2021} has proposed to start simulations as late as possible from high-order LPT in order to suppress discreteness errors. Here, we follow a distinctly different approach: we improve the simulations, both at the level of the time integration, and at the level of the force computation, in order to demonstrate agreement. In the following sections, we detail the steps that are necessary to achieve this.

\subsection{\texorpdfstring{$\boldsymbol{N}$-body}{N-body} time integrators -- symplectic and/or fast}

\noindent In this section, we will provide some basics on LPT-inspired integrators, focusing on relevant details for the \textsc{PowerFrog} integrator recently introduced in Ref.~\cite{list2023perturbation}.
As discussed above, LPT expands the displacement field in Einstein-de Sitter cosmology in a Taylor series in terms of scale factor time $a$ in the pre-shell-crossing regime. This ansatz is readily carried over to an analogous series in terms of the linear growth-factor time $D$ for $\Lambda$CDM cosmology (neglecting higher-order correction terms derived in Ref.~\cite{Rampf:2022,2022PhRvD.106l3504F}). In particular, in one dimension, all terms of the series with order $n \geq 2$ vanish, and only the Zel'dovich solution remains, i.e., $\vecb{\mPsi}(\vecb{q}, D) = -D \, \bnabla_{\vecb{q}} \phi_{\text{ini}}$.

This suggests that employing the momentum w.r.t.\ $D$-time, $\vecb{\mPi} = \dd \vecb{X} / \dd D = \vecb{P} / F$, for the time integration of cosmological $N$-body systems allows matching the dynamics on large scales to the Zel'dovich approximation or even higher LPT orders. Here, $\vecb{X}$ and $\vecb{P} = H(a) \, a^3 \, \dd \vecb{X} / \dd a$ denote the canonical position and momentum variable, respectively, and the factor relating $\vecb{\mPi}$ and $\vecb{P}$ is given by $F(a) = H(a) \, a^3 \, \dd D / \dd a$ with the Hubble parameter $H(a)$. 

In Ref.~\cite{list2023perturbation}, we introduced the so-called $\mPi$-integrators which implement this idea and advance the (noncanonical) position and momentum pair $(\vecb{X}_i, \vecb{\mPi}_i)$ for each particle $i$ from time step $k$ to $k + 1$ according to
\begin{subequations}
\begin{align}
    \vecb{X}_i^{k+\nicefrac{1}{2}} &= \vecb{X}_i^{k} + \frac{\Delta D}{2} \vecb{\mPi}^{k}_i, \label{eq:drift_1} \\
    \vecb{\mPi}_i^{k+1} &= p(D_k, D_{k+1}) \, \vecb{\mPi}_i^{k} - q(D_k, D_{k+1}) \, a_{k + \nicefrac{1}{2}} \, \bnabla_{\vecb{x}} \varphi_N(\vecb{X}^{k+\nicefrac{1}{2}})_i, \label{eq:kick} \\
    \vecb{X}_i^{k+1} &= \vecb{X}_i^{k+\nicefrac{1}{2}} + \frac{\Delta D}{2} \vecb{\mPi}^{k+1}_i, \label{eq:drift}
\end{align}
\end{subequations}
where $\varphi_N(\vecb{X})$ is the gravitational potential induced by the $N$ simulation particles located at positions $\vecb{X}$ via Poisson's equation $\nabla^2_{\vecb{x}} \varphi_N = 3/(2a) \, H_0^2 \, \Omega_m \, \delta_N$.
Here, we defined $\Delta D = D_{k+1} - D_{k}$ as the length of the time step w.r.t.\ growth-factor time.
For the kick, the coefficients $p$ and $q$ can in principle be chosen independently; however, we showed in Ref.~\cite{list2023perturbation} that a $\mPi$-integrator reproduces the exact Zel'dovich solution in 1D until shell-crossing in a single time step if and only if $p$ and $q$ are related via 
\begin{equation}
    1 - p = \frac{3}{2} \Omega_m \, H_0^2 \, D_{k + \nicefrac{1}{2}} \, q.
    \label{eq:zeldovich_consistency}
\end{equation}
Also, we showed that the only \textit{symplectic} integrator that satisfies the aforementioned relation is the well-known \textsc{FastPM} integrator by Ref.~\cite{Feng:2016}, which corresponds to the choice $p(D_n, D_{n+1}) = F(D_n) / F(D_{n+1})$. However, in view of the expansion of the universe, it is questionable whether symplecticity is a necessary property for time integrators when considering large scales where the particle motion is largely governed by the Hubble flow. Indeed, we introduced new (non-symplectic) integrators in Ref.~\cite{list2023perturbation}, which perform better than \textsc{FastPM} in terms of the power spectrum and cross-spectrum for any given number of time steps. One of these integrators, which we named \textsc{PowerFrog}, is explicitly constructed to match the 2LPT asymptote at early times $a \to 0$. For a single time step starting from $D(a = 0) = 0$ to some final growth-factor time $\Delta D$, the coefficient functions $p(0, \Delta D)$ and $q(0, \Delta D)$ of \textsc{PowerFrog} are simply given by
\begin{equation}
    p(0, \Delta D) = -\frac{5}{7}, \qquad q(0, \Delta D) = \frac{16}{7 \, \Omega_m \, H_0^2 \, \Delta D}.
\end{equation}
For later times $D_k > 0$, the function $p(D_k, D_{k+1})$ in the kick takes the form
\begin{equation}
    p(D_k, D_{k+1}) = \frac{\alpha D_k^\epsilon + \beta D_{k+1}^\epsilon}{D_{k+\nicefrac{1}{2}}^\epsilon} + \gamma,
\end{equation}
where the coefficients $\alpha, \beta, \gamma, \epsilon$ need to be determined numerically as the solution of a transcendental system of equations that ensures (global) second-order convergence (eliminating three degrees of freedom) and consistency with the 2LPT asymptote (eliminating the fourth degree of freedom), see Ref.~\cite{list2023perturbation}. The coefficient $q$ then follows from Eq.~\eqref{eq:zeldovich_consistency}.

\section{Particle discreteness reduction in cosmological particle-in-cell codes}

\noindent The main limitation in the convergence of $N$-body simulations to the fluid limit is due to the finite sampling by particles. In simulations, the continuous indexing by the Lagrangian coordinate $\vecb{q}\in\mathbb{T}^3$ is replaced by a discrete collection of characteristics, which we initially arrange in terms of a simple cubic lattice $\vecb{q}_{\vecb{i}}:=\vecb{i}/\mathcal{N}$ where $\vecb{i}\in\mathbb{I}_{\mathcal{N}} := (\mathbb{Z}/\mathcal{N} \mathbb{Z})^3$. Here, $\mathcal{N}$ is the linear number of particles, i.e.\ $\mathcal{N}^3 = N$.  In particular, we then have a discrete set of particle trajectories $a\mapsto \vecb{X}_{\vecb{i}}(a) := \vecb{X}(\vecb{q}_{\vecb{i}},a)$ along with their displacement vectors $\vecb{\mPsi}_{\vecb{i}}(a) := \vecb{X}_{\vecb{i}}(a)-\vecb{q}_{\vecb{i}}$. We distinguish between active `characteristic' particles and passive `mass' carrying particles. In standard particle-in-cell (PIC) / PM simulations, these two particles are identical. We shall clarify this distinction in what follows.

\subsection{Spectral sheet interpolation}
\begin{figure}
  \includegraphics[width=.9\textwidth]{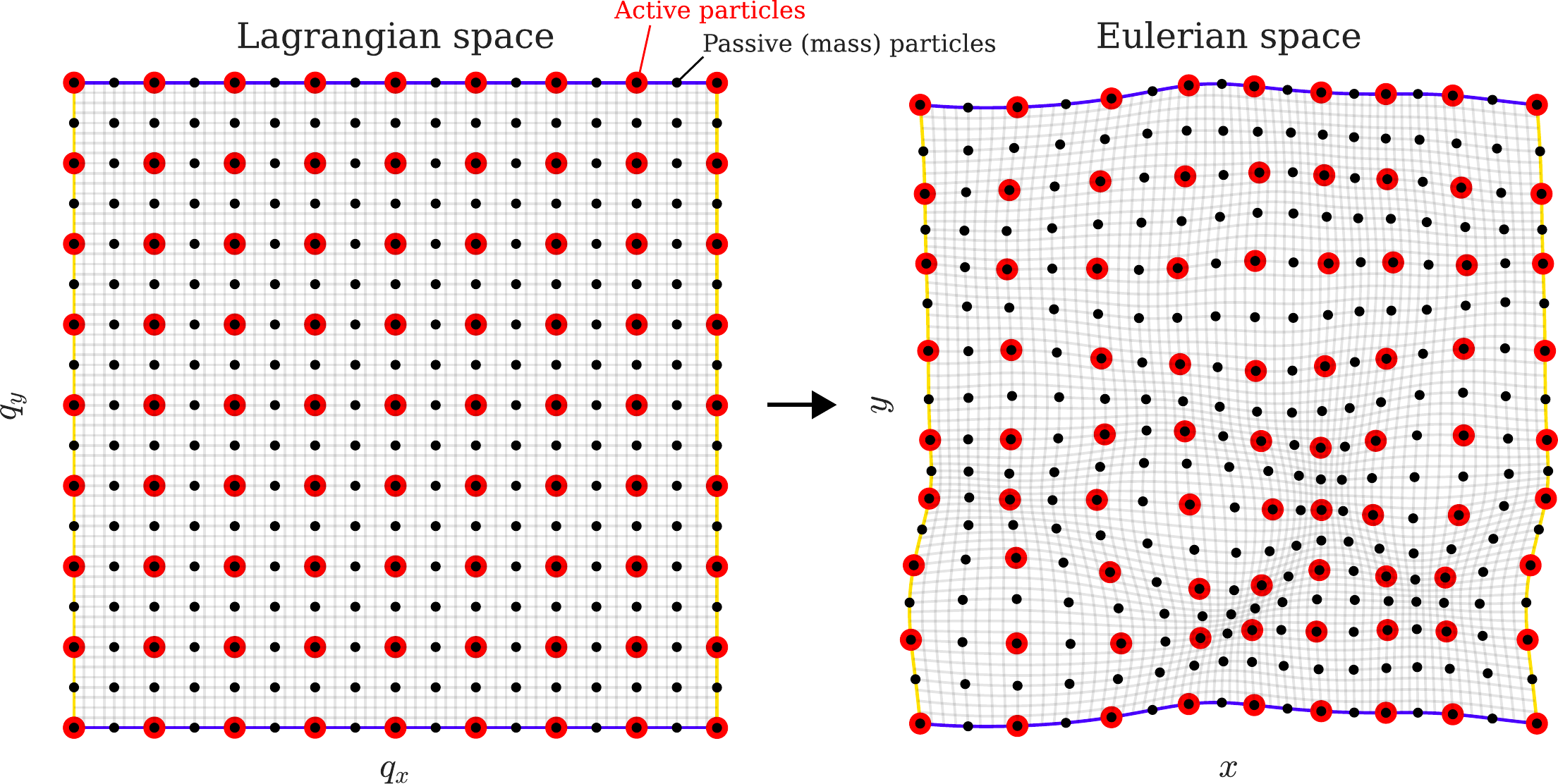}
  \caption{\label{fig:resampling} Two-dimensional sketch of the spectral sheet interpolation, which reduces the discreteness of the potential when computing the acceleration. 
  The particle distribution is upsampled by a factor of $R = 2$ per dimension in this illustration. Red dots represent the $N$ \textit{active} particles, whose trajectories are traced by the $N$-body simulation. Black dots are \textit{passive} mass particles, which are spawned on a finer uniform grid in Lagrangian space (\textit{left panel}), and their resulting Eulerian positions (\textit{right}) are determined by Fourier interpolating the displacement field. Note that a passive particle is located at each active particle, and for $R = 1$ (i.e.\ no upsampling), the sets of active and passive particles are identical. The yellow and blue domain boundaries in each panel coincide due to the torus topology of the simulation box. The faint gray grid lines illustrate the sheet interpolation with a higher upsampling factor of $R = 8$: in that case, passive particles would be located at all grid line intersections.}
\end{figure}

\noindent Due to its cold nature, CDM occupies only a three-dimensional submanifold of phase space. This property can be exploited to interpolate the displacement field to new mass resolution elements \cite{Abel:2012,Shandarin:2012} in order to approach the continuum limit. This has already been used for `sheet-based' simulations \cite{Hahn:2013,Hahn:2016,Sousbie:2016} that are known to overcome some of the well-known discreteness problems of gravitational $N$-body simulations, such as artificial fragmentation (e.g.~\cite{10.1111/j.1365-2966.2007.12053.x, Romeo_2008}). While previous work in this direction has mostly employed low-order interpolation on the Lagrangian submanifold (tetrahedral \cite{Abel:2012, Shandarin:2012,Hahn:2013,Sousbie:2016}, tri-linear, tri-quadratic \cite{Hahn:2016}), here we use Fourier interpolation to achieve spectral accuracy (see also \cite{stucker2018median}). To this end, we define the Fourier-space translation operator
\begin{align}
  \hat{T}_{\vecb{y}}f(\vecb{x}) := f(\vecb{x}+\vecb{y}) = \mathcal{F}^{-1}\left[ \ee^{\ii \vecb{k}\cdot \vecb{y}}\mathcal{F}[f]\right](\vecb{x}),
\end{align}
implemented with a discrete Fourier transform. Given a Lagrangian shift vector $\vecb{s}$ and our set of active characteristic particles, we can then generate new sets of sheet-interpolated `mass' particles by evaluating
\begin{align}
  \vecb{X}^{\vecb{s}}(\vecb{q}_{\vecb{i}},a) := \vecb{q}_{\vecb{i}}+\vecb{s} + \hat{T}_{\vecb{s}}\mPsi(\vecb{q}_i,a).
\end{align}
As the particles are occupying a simple cubic lattice, we have the invariance $\vecb{X}^{\vecb{i}'/\mathcal{N}}(\vecb{q}_{\vecb{i}},a) = \vecb{X}(\vecb{q}_{\vecb{i}'-\vecb{i}},a)$. By choosing subdivisions of the unit cube, we can upsample the particle distribution $R$ times per dimension, i.e.\ $\vecb{s} = \vecb{m}/(\mathcal{N} R)$ where $\vecb{m}\in [0, 1, \dots, R-1]^3$. An illustration of the sheet-based upsampling is shown in Fig.~\ref{fig:resampling}.

\subsection{Mass assignment / interpolation kernels}

\noindent In PIC/PM codes, the initial distribution function $f_{\text{ini}}(\vecb{x},\vecb{p})$ is sampled by the $N$ active characteristic particles. As described above, we allow for upsampling them to a larger number of $N_{\text{mass}} \ge N$ passive mass particles,
\begin{align}
  \hat{n}(\vecb{x},a) := \frac{1}{N_{\text{mass}}} \sum_{j=1}^{N_{\text{mass}}} \delta_{\rm D}(\vecb{x}-\vecb{X}_j(a)).
\end{align} 
The particle mesh is given by the 3-dimensional Dirac comb of uniform spacing $h>0$, i.e.\ the object
  \begin{align}
    \Sha_h(\vecb{x}):=\sum_{\vecb{n}\in\mathbb{Z}^3}\delta_{\rm D}(\vecb{x}-h\vecb{n}).
  \end{align}
Given a mass assignment kernel $W(\vecb{x})$, the grid-interpolated particle distribution can be written as \cite{HockneyEastwood}
\begin{align}
  \mathfrak{n}(\vecb{x},a) := \Sha_h(\vecb{x}) \,\left(\hat{n}(\vecb{x},a) \ast W(\vecb{x})\right), \label{eq:mass_PM}
\end{align}
where the asterisk denotes convolution.
Kernels of order $n$ are generated by $n$ convolutions of the box function with itself, i.e.\
\begin{subequations}
\begin{align}
  W_{\rm NGP}(x) := W_{1}(x)&= \frac{1}{h}\left\{ \begin{array}{ll} 
    1 & \quad \textrm{for }\left| x \right| \le \frac{h}{2}\\
    0 & \quad \textrm{otherwise}
    \end{array}\right. \\
    W_{\rm CIC}(x) := W_{2}(x) &= \frac{1}{h}\left\{ \begin{array}{ll} 
    1-\frac{\left| x\right|}{h} &\quad  \textrm{for }\left|x\right| < h \\
    0 &\quad  \textrm{otherwise}
    \end{array}\right. \label{eq:cic_kernel}\\
    W_{\rm TSC}(x) := W_{3}(x) &= \frac{1}{h}\left\{ \begin{array}{ll} 
    \frac{3}{4} - \left( \frac{x}{h}\right)^2 & \quad \textrm{for }\left|x\right|\le \frac{h}{2}\\
    \frac{1}{2}\left( \frac{3}{2} - \frac{\left|x\right|}{h}\right)^2 & \quad \textrm{for }\frac{h}{2}\le \left| x \right| < \frac{3h}{2}\\
    0 & \quad \textrm{otherwise}
    \end{array}\right. \\
    W_{\rm PCS}(x) := W_{4}(x) &= \frac{1}{h}\left\{ \begin{array}{ll} 
    \frac{1}{6} \left[ 4 - 6\left(\frac{x}{h}\right)^2  + 3 \left(\frac{|x|}{h}\right)^3 \right] & \quad \textrm{for }\left|x\right|\le h\\
    \frac{1}{6}\left( 2 - \frac{\left|x\right|}{h}\right)^3 & \quad \textrm{for } h \le |x| < 2 h\\
    0 & \quad \textrm{otherwise}
    \end{array}\right.
\end{align}
where the three-dimensional version is simply given by the 
product of three one-dimensional kernel evaluations. Although most cosmological simulations use $n=2$ (CIC), we used $n=4$ (PCS) to sufficiently reduce particle discreteness effects for our precision study. Ref.~\cite{Chaniotis:2004} lists kernels of even higher order, which we did, however, not use. They have the Fourier transform 
\begin{align}
  \widetilde{W}_{n}(k) = \left[{\rm sinc}(hk / 2)\right]^n.
\end{align}
\end{subequations}

\subsection{Poisson solver}

\noindent We use an FFT-based spectral Poisson solver. Given the discretized mass distribution on the grid from Eq.~\eqref{eq:mass_PM}, we obtain the acceleration field as 
\begin{align}
  \vecb{\mathfrak{a}}(\vecb{x}) := \mathcal{F}^{-1}\left[ -\frac{\ii \vecb{k}}{\|\vecb{k}\|^2}\; \widetilde{W}_n^{-2}(\vecb{k})\;\;\mathcal{F}[\mathfrak{n}](\vecb{k})\right] \,, \label{eq:acceleration}
\end{align}
where $\mathfrak{n}$ is defined in Eq.~\eqref{eq:mass_PM}.
The acceleration field is then interpolated back to the active particle positions using once more the mass assignment kernel $W$. The double deconvolution with $W$ accounts for both the deposit and the back-interpolation to the active particles. 

Instead of solving the Poisson equation in Fourier space (where the discretization occurs at the level of the FFT, which only takes into account a finite number of modes), it is also common practice to compute the derivatives in real space using finite-difference (FD) approximations. For instance, the popular \textsc{Gadget-2} code \cite{Gadget2} uses a fourth-order stencil for the gradient whose 
Fourier transform is given by
\begin{equation}
    \mathcal{F}[\partial_{x_d}](\vecb{k}) = \frac{\ii}{6} \left(8 \sin(k_d) - \sin(2 k_d)\right), \qquad \text{for } d \in \{1, 2, 3\},
\end{equation}
where $\vecb{k} = (k_1, k_2, k_3)^\top$ is in units of the grid's Nyquist wavenumber.
In our ablation study, we analyze the effect of replacing the spectral $\ii \vecb{k}$ gradient kernel by this commonly employed fourth-order FD approximation; however, we still perform the differentiation in Fourier space (as done in e.g.\ Ref.~\cite{Feng:2016}).
Note that due to the odd symmetry of FD stencils for the gradient, their Fourier transform is given by the sum of sines, which must vanish at the Nyquist wavenumber (see the discussion in Ref.~\cite[Sec.~3.4]{Hahn2011}). The FD gradient operators therefore effectively act as low-pass filters, which suppress power close to the Nyquist frequency.

While the situation is, in principle, similar for the Laplacian operator (with FD approximations giving rise to a truncated cosine series in Fourier space), we noticed very little difference when replacing the spectral Laplacian $-\|\vecb{k}\|^2$ by FD operators, for which reason we do not include any ablation tests for the Laplacian herein.

\subsection{Dealiasing by interlacing}
\begin{figure}
  \includegraphics[width=0.8\textwidth]{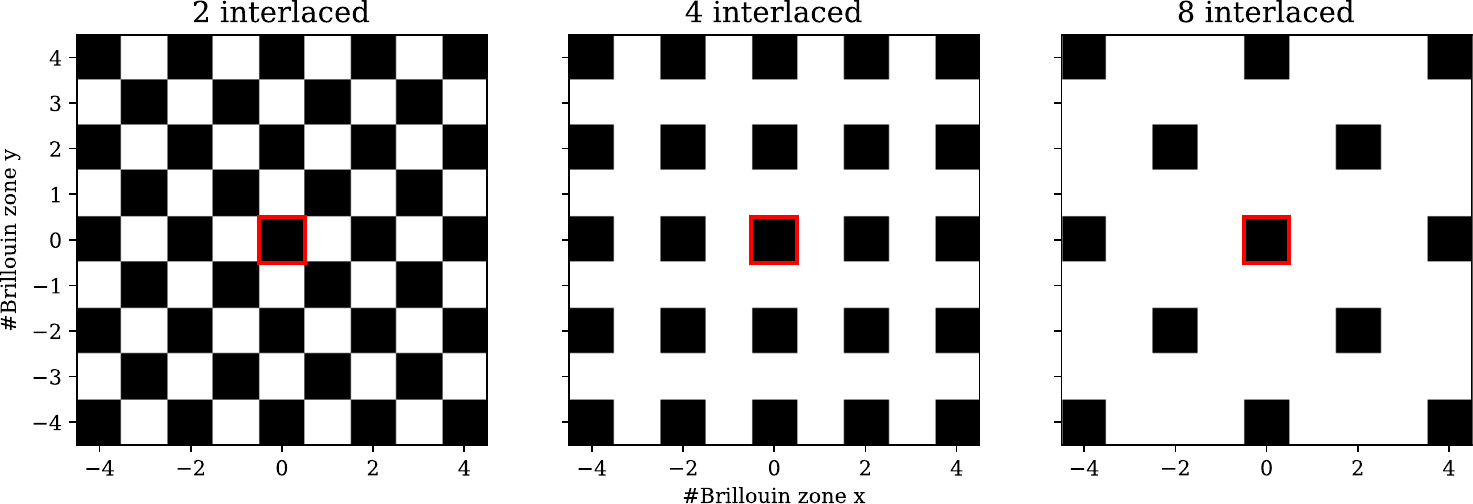}
  \caption{\label{fig:aliasing} Removal of aliases with increasingly higher-order interlacing schemes. While our simulations are three-dimensional, the procedure is shown here for two dimensions for illustrative purposes. The left panel illustrates the interlacing of two grids \cite{HockneyEastwood,sefusatti2016accurate}---shifted by $(\nicefrac{h}{2},\nicefrac{h}{2})$ w.r.t.\ each other---which removes every second alias and therefore the dominant aliasing contribution (removed aliases are white). This procedure can be continued to four interlaced grid removing increasingly more aliases by including also the shifts $(\nicefrac{h}{2},0)$ and $(0,\nicefrac{h}{2})$ (middle panel), or to eight interlaced grids, where further $h(\nicefrac{1}{2}\pm\nicefrac{1}{4},\nicefrac{1}{2}\pm\nicefrac{1}{4})$ are included. The red cell indicates the first Brillouin zone at $(0,0)$ in each case.}
\end{figure}

\noindent PIC/PM simulations suffer from aliasing since the particle distribution is not band limited. Aliasing could in principle be reduced by increasing the sampling rate, i.e.\ resolution of the grid, which is however prohibitive due to its memory requirements. It is well known that interlacing techniques can be used to eliminate the dominant aliases and effectively achieve sampling comparable to higher resolution. Here, we adopt the technique proposed by Ref.~\cite{Chen:1974} (see also \cite{HockneyEastwood, sefusatti2016accurate}) and use interlaced grids in order to remove the dominant aliasing contributions in the accelerations.

It is easy to show (e.g. \cite[Sec.~7.8]{HockneyEastwood}) that by depositing the particles onto a grid $\Sha_h^{(\nicefrac{1}{2}, \nicefrac{1}{2}, \nicefrac{1}{2})}$ that is shifted by half a grid-cell w.r.t.\ the original grid $\Sha_h$, i.e.
\begin{equation}
    \Sha_h^{(\nicefrac{1}{2}, \nicefrac{1}{2}, \nicefrac{1}{2})}(\vecb{x}) := \sum_{\vecb{n}\in\mathbb{Z}^3}\delta_{\rm D}\left(\vecb{x} - h \left(\vecb{n} + (\nicefrac{1}{2}, \nicefrac{1}{2}, \nicefrac{1}{2})^\top \right)\right),
\end{equation}
and averaging the two accelerations resulting from the grids $\Sha_h$ and $\Sha_h^{(\nicefrac{1}{2}, \nicefrac{1}{2}, \nicefrac{1}{2})}$, half of the aliases can be removed (namely those for which $n_1 + n_2 + n_3 = \text{odd}$, where $\vecb{n} = (n_1, n_2, n_3)^\top$ now indexes the reciprocal lattice). The resulting checkerboard pattern on the reciprocal lattice is illustrated in the first panel of Fig.~\ref{fig:aliasing} (in 2D for illustrative purposes), with aliased (dealiased) Brillouin zones shown in black (white). By extending this idea to more than two shifted grids, higher-order alias contributions can be removed, see the second and third panel in Fig.~\ref{fig:aliasing}.

In practice, given a set of shift vectors $\mathcal{D} = \{\vecb{d}_1, \ldots, \}$ with $\vecb{d}_i \in [0, 1)^3$ for $i \in \{1, \ldots, D\}$, we implement the interlacing as follows:

\begin{algorithm}[H]
\For{$\vecb{d} \in \mathcal{D}$}{
    Compute the (resampled) overdensity according to Eq.~\eqref{eq:mass_PM} on the shifted grid $\Sha_h^{\vecb{d}}(\vecb{x})$\;
    Solve the Poisson equation and obtain the acceleration field $\vecb{\mathfrak{a}}(\vecb{x})$ on the grid via Eq.~\eqref{eq:acceleration}\;
    Interpolate the accelerations back to the particles, using the same mass assignment kernel $W(\vecb{x})$\;
}
Average the accelerations over all interlaced grids $\mathcal{D}$\;
\label{alg:interlacing}
\end{algorithm}

\noindent Of course, it is not necessary to store the accelerations for each grid in memory, but one can simply add the new to the currently stored accelerations in each loop iteration and divide the final result by $D$ to obtain the dealiased acceleration field.

\subsection{Non-uniform Fast Fourier Transform}

\noindent As explained above, the density in PM simulations is computed by convolving the discrete point particles with a (localized) mass assignment kernel in order to obtain the grid-interpolated particle density $\mathfrak{n}$ as defined in Eq.~\eqref{eq:mass_PM}. Then, the density field is Fourier transformed, and the Poisson equation is solved in Fourier space, see Eq.~\eqref{eq:acceleration}.

In order to keep the notation simple, let us consider the one-dimensional case in this section, noting that the three-dimensional DFT simply performs a one-dimensional DFT along each of the three axes. Recall that the discrete Fourier transform (DFT) maps a sequence $(c_m)_{m=1}^M$ of $M$ numbers (which are real in our case and correspond to the grid-interpolated particle density $\mathfrak{n}$ at each of the $M$ grid points) to a sequence of complex numbers $(C_m)_{m=1}^M$ via
\begin{equation}
    C_m = \sum_{\ell = 1}^{M} c_\ell {\rm e}^{-2 \pi \ii x_\ell m},
    \label{eq:dft}
\end{equation}
where $x_\ell = \ell / M$ are the grid points. However, loosening the restriction that the points must lie on a uniform grid, and further allowing the length of the input sequence to be the number of \textit{particles} $N$, rather than the number of \textit{grid points} $M$, i.e.\ $(c_n)_{n=1}^N$ (while allowing for a potentially different number of considered Fourier modes $M \neq N$), one could evaluate the sum in Eq.~\eqref{eq:dft} directly using the exact \textit{particle} locations $x_n$ (rather than the grid points), with factors $c_n \equiv 1$, as we take all particles to have unity mass, i.e.\
\begin{equation}
    C_m = \sum_{n=1}^{N} {\rm e}^{-2 \pi \ii x_n m},
    \label{eq:nudft}
\end{equation}
without the need for a localized mass assignment kernel $W$.
Exactly this is accomplished by the so-called \textit{non-uniform DFT}, which computes Eq.~\eqref{eq:nudft} for \textit{arbitrary} positions $x_n$ such as the (Eulerian) particle positions in our case. Since the explicit interpolation step using CIC, TSC, etc.\ onto a mesh is not required in this case, computing the Fourier density directly based on the particle locations turns out to be another effective way of reducing discreteness effects. For instance,dealiasing via interlaced grids as described above is not necessary when using the non-uniform FFT, simply because there is no mesh. 

In practice, we use the non-uniform FFT wrapper by Ref.~\cite{jaxfinufft} around \textsc{cuFinufft}, which is based on Refs.~\cite{barnett2019parallel, barnett2021aliasing, shih2021cufinufft}. With this algorithm, the particle positions are also interpolated onto a mesh, and a `standard' FFT is performed on the mesh; however, the deviation of the result from the exact Fourier sum in Eq.~\eqref{eq:nudft} can be explicitly controlled, with a higher desired accuracy leading to a larger support of the interpolation kernel. From this point of view, the non-uniform FFT leads to a force computation similar to its usual PM-based counterpart with CIC, TSC, etc.\ interpolation, but with the difference that the kernel is chosen large enough to guarantee spectral accuracy to a specified degree. While we believe that it would be interesting to explore the non-uniform FFT more generally in the context of cosmology (where the computation of Fourier-based statistics such as the power spectrum of irregular data is ubiquitous) and specifically for the force computation in cosmological simulations, we defer a detailed investigation of aspects such as run time and memory requirements to future work.

\subsection{Tree-PM}

\noindent As an additional test, we also consider the force computation via the Tree-PM method \cite{bagla2002treepm, Bode2003}. In this approach, the gravitational force is split into a long-range force, computed with the PM method, and a short-range force, computed with the Barnes--Hut tree method \cite{appel1985efficient, barnes1986hierarchical}. 
The gravitational potential is written as the sum of a short-range (S) and a long-range (L) part
\begin{equation}
    \varphi = \varphi^{(\text{S})} + \varphi^{(\text{L})},
\end{equation}
giving rise to the Fourier-space Poisson equation
\begin{equation}
    \hat{\varphi}^{(\text{L})}(\vecb{k}) = - \frac{3}{2a \|\vecb{k}\|^2} H_0^2 \Omega_m \hat{\delta} \exp\left(-\frac{1}{2} \|\vecb{k}\|^2 r_\alpha^2 \right)
\end{equation}
for the long-range part (which is the same as for the PM method, except for an exponential cut-off for large wave numbers as determined by the force-split scale $r_\alpha$), and the short-range part, which is computed in real space and reads as

\begin{equation}    
        \varphi^{(\text{S})}(\vecb{x}) = - \frac{3}{2a} \frac{H_0^2 \Omega_m}{4 \pi} \sum_{i = 1}^N \frac{1}{\|\vecb{x} - \vecb{X}_i\|} \operatorname{erfc}\left(\frac{\|\vecb{x} - \vecb{X}_i\|}{\sqrt{2} \, r_\alpha}\right).
\end{equation}
The tree part neglects the infinitely many periodic copies of each particles and only takes into account the nearest gravity source particle (w.r.t.\ the 3-torus topology). 
Also, we note that the tree part of the force computation is currently performed on a CPU, not on a GPU.

\subsection{Specific choices of the discreteness reduction parameters}

\noindent Having explained the different discreteness suppression techniques, we can now summarize the specific settings we used when performing a \textsc{PowerFrog} step from $z = \infty$ to $z = 18$ in order to achieve the small residuals towards $3 - 4$LPT (see Fig.~\ref{fig:displacement_plot}). Also, the discreteness-suppressed ICs for our Gadget runs from $z = 36$ to $z = 0$ were generated with the same parameters (i.e.\ the ICs for all curves with a label ``Suppression'' in Fig.~\ref{fig:pk_plot_z_0}).

\begin{itemize}
    \item Spectral sheet interpolation for spawning $2^3$ `mass' particles from each `characteristic' particle (i.e.\ $R = 2$), based on which the density field is computed
    \item PCS mass assignment
    \item Dealiasing by using $D = 2$ interlaced grids (the resulting Brillouin pattern is the 3D counterpart of the one for the `2 interlaced' case in 2D shown in Fig.~\ref{fig:aliasing})
    \item Exact spectral gradient kernel ($\ii \vecb{k}$) and double deconvolution with the MAK when solving Poisson's equation
\end{itemize}

For the non-uniform FFT results, we also used $2^3$ mass particles ($R = 2$), and we used $M = (7/4)^3 N = 896^3$ Fourier modes (due to memory issues with $M = 2^3 N$), which turned out to be sufficient to achieve the same discreteness suppression as with our PM baseline with $M = 2^3 N = 1024^3$ PM grid cells (and hence Fourier modes). In our Tree-PM implementation, we took $r_\alpha$ to be $1.5\times$ the cell size of the PM mesh such that force anisotropies due to the orientation of the PM mesh are suppressed (e.g.~\cite{Gadget2}).

Note that for Fig.~\ref{fig:resampling-plot} which illustrates the effect of the resampling (and to a lesser extent of the other discreteness reduction techniques), we used $5^3$-fold resampling, i.e.\ $R = 5$.

\section{Additional material for the discreteness-suppressed \textsc{PowerFrog} step from \texorpdfstring{$\boldsymbol{z = \infty}$}{redshift infinity}}

\begin{figure}
    \centering
    \resizebox{1\textwidth}{!}{
    \includegraphics{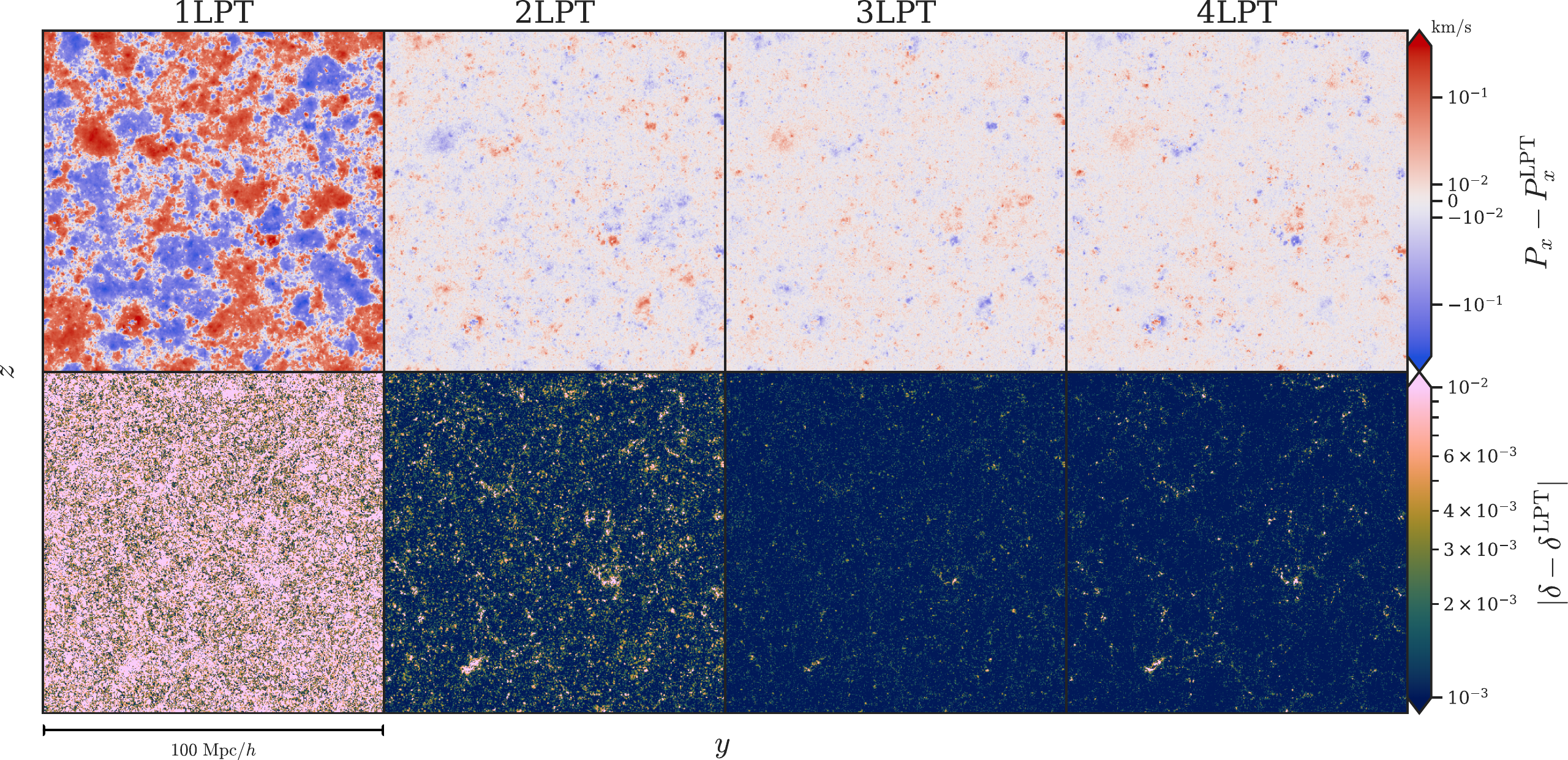}
    }
    \caption{Residuals of the Lagrangian (comoving) $x$-velocity $P_x$ (\textit{top}) and the density contrast $\delta$ (\textit{bottom}) between our results with a single \textsc{PowerFrog} step from $z = \infty$ to $z = 18$ and the corresponding LPT fields at $z = 18$ for different LPT orders (see Fig.~\ref{fig:displacement_plot} in the main body for the displacement residual). Shown is a single slice in the $y$-$z$ plane.}    
    \label{fig:P_delta_plot}
\end{figure}

\noindent For completeness, we show the velocity and density residuals after a single \textsc{PowerFrog} step from $z = \infty$ to $z = 18$ w.r.t.\ different LPT orders in Fig.~\ref{fig:P_delta_plot} (cf.\ Fig.~\ref{fig:displacement_plot} in the main body for the residual of the displacement field). Since the displacement field lies much closer to $3-4$LPT than to 2LPT, the same is true for the density field. In contrast, the velocity residual towards 3LPT is only slightly smaller than towards 2LPT.  This might be related to the fact that the third-order term after the \textsc{PowerFrog} step matches 3LPT more closely for the displacements than for the velocities; we will analyze this in depth in future work.

\section{Additional material for the ablation study}
\subsection{Displacement residuals at the field level}

\begin{figure*}
    \centering
    \resizebox{0.8\textwidth}{!}{
    \includegraphics{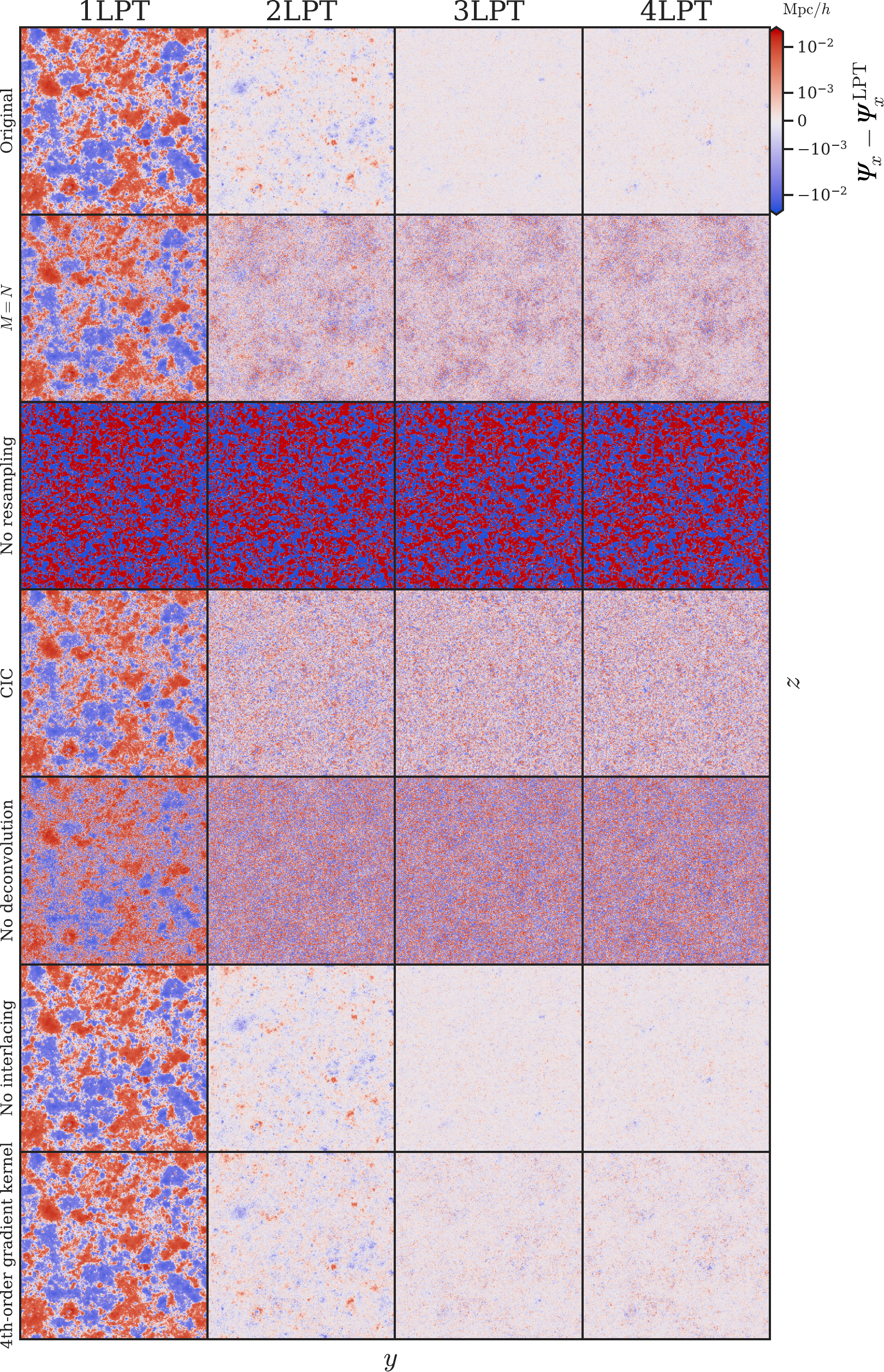}
    }
    \caption{Residuals of the $x$-displacement $\mPsi_x$ between our results with a single step from $z = \infty$ to $z = 18$ (with PM force computation) and the corresponding LPT fields at $z = 18$ for different LPT orders (\textit{columns}). The first row shows the baseline case (same as Fig.~\ref{fig:displacement_plot} in the main body). In the following rows, one discreteness suppression technique at a time is omitted. Shown is a single slice in the $y$-$z$ plane.}
    \label{fig:displacement_plot_ablation}
\end{figure*}

\begin{figure*}
    \centering
    \resizebox{1\textwidth}{!}{
    \includegraphics{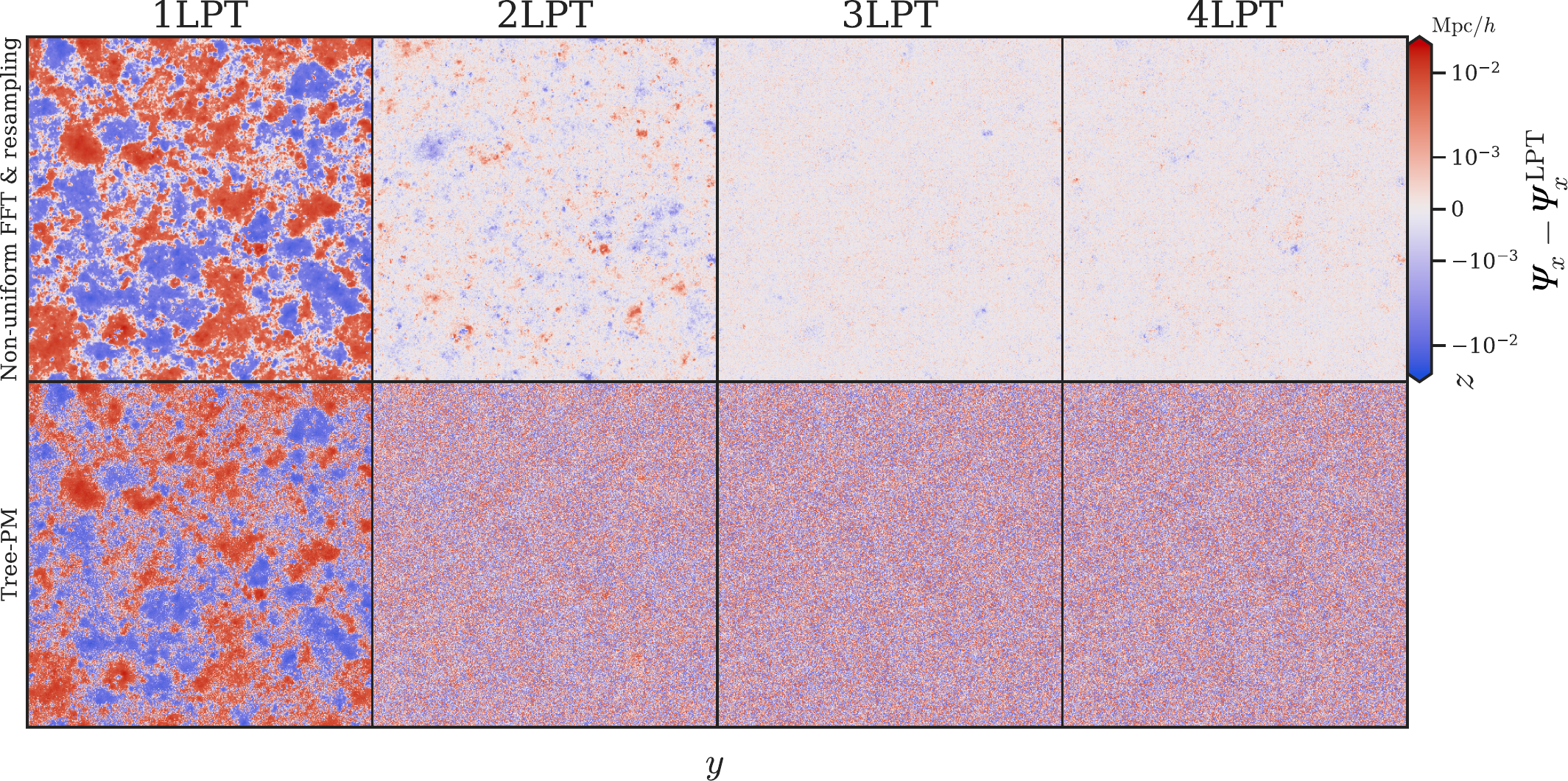}
    }
    \caption{Same as Fig.~\ref{fig:displacement_plot} in the main body, but when replacing the PM force computation by the non-uniform FFT with resampling (\textit{top}) and with Tree-PM (without resampling, \textit{bottom}). The non-uniform FFT with resampling achieves the same noise reduction as our PM baseline (see also Fig.~\ref{fig:displacement_residual_rms}).}
    \label{fig:displacement_plot_ablation2}
\end{figure*}

\noindent In Fig.~\ref{fig:psi_rms_plot} in the main body of this work, we studied the effect of the different discreteness reduction techniques on the relative RMS error between the single-step displacement field at $z = 18$ and different LPT orders. To provide a more intuitive understanding of these errors, we plot the displacement field residuals for the different cases in Figs.~\ref{fig:displacement_plot_ablation} and~\ref{fig:displacement_plot_ablation2}.

The first row shows again the residual of the displacement fields between a single $N$-body step from $z = \infty$ to $z = 18$ with the \textsc{PowerFrog} stepper, using all discreteness suppression methods discussed above (i.e.\ the same results as Fig.~\ref{fig:displacement_plot}). Each subsequent row depicts the results when omitting one of these techniques at a time. As discussed in the main text, the sheet-based resampling has by far the biggest impact, followed by the deconvolution of the mass assignment kernel, using PCS instead of CIC, and taking a PM grid at twice the particle resolution, i.e.\ $M = 2^3 N$, rather than $M = N$. The other techniques have a smaller effect; however, each of them contributes to reducing the discreteness noise in the `Original' row, where the $3-4$LPT residual is clearly dominated by patch-like structures, rather than high-frequent noise. These patches stem from the fact that---as expected---the single \textsc{PowerFrog} step does not entirely capture the $3-4$LPT terms, for which reason they cannot be removed by suppressing discreteness even further.

\begin{figure*}
    \centering
    \resizebox{\textwidth}{!}{
    \includegraphics{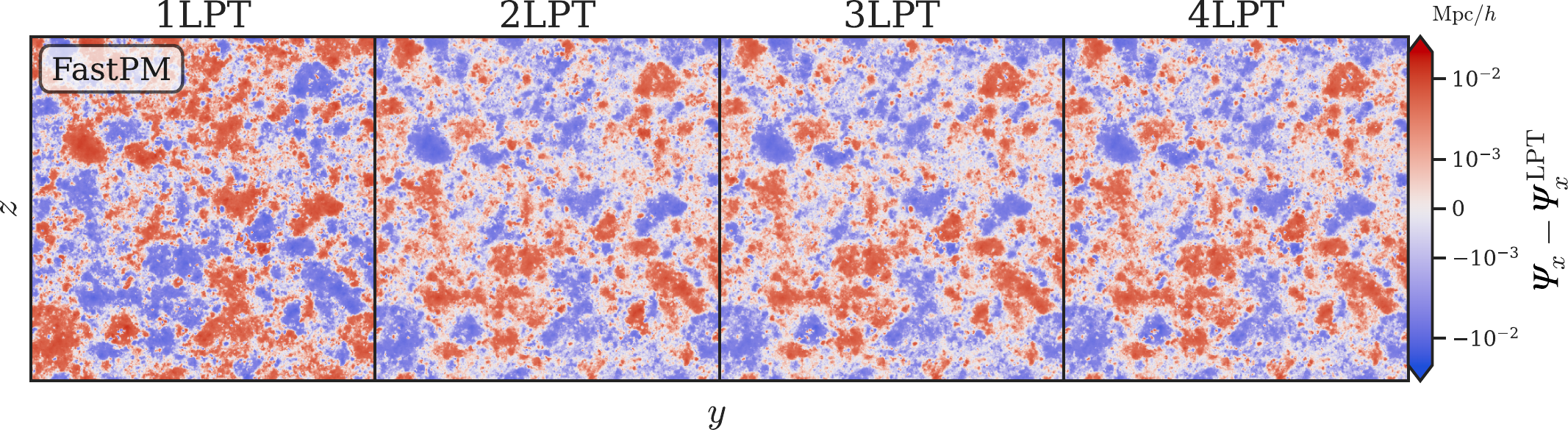}
    }
    \caption{Same as Fig.~\ref{fig:displacement_plot} in the main body, but with a single DKD step of the \textsc{FastPM} time integrator rather than \textsc{PowerFrog}. Evidently, a single \textsc{FastPM} step does not correctly capture the 2LPT term.}
    \label{fig:displacement_plot_fastpm}
\end{figure*}

We also repeat the 1-step simulation from $z = \infty$ with the popular \textsc{FastPM} integrator \cite{Feng:2016}, which correctly reproduces the 1LPT (i.e.\ Zel'dovich) growth, but has asymptotics different from 2LPT for $z \to \infty$. While \textsc{FastPM} is often employed in kick-drift-kick form, we perform a single drift-kick-drift step here, noting that the acceleration at $z = \infty$ vanishes for particles placed on a homogeneous grid, for which reason starting with a kick at $z = \infty$ would be futile. Figure \ref{fig:displacement_plot_fastpm} shows a slice of the residual between the 1-step \textsc{FastPM} simulation and different LPT orders (cf. Fig.~\ref{fig:displacement_plot} in the main body for the same plot with the \textsc{PowerFrog} stepper). Clearly, there is a significant 2LPT contribution in the residual, which is also present in the residuals w.r.t.\ higher LPT orders (see Fig.~\ref{fig:psi_rms_plot} for a quantitative assessment). \textsc{FastPM} should therefore not be used for initializing cosmological simulations in a single step from $z = \infty$, just as any other integrator that is merely `Zel'dovich-consistent' in the sense of Def.~3 in Ref.~\cite{list2023perturbation}. We remark, however, that \textsc{PowerFrog} is not the only possible choice for an integrator with correct 2LPT asymptotics, and one can in principle construct integrators whose behavior for $z \to \infty$ explicitly matches higher LPT orders $> 2$.

\begin{figure*}
    \centering
    \resizebox{0.6\textwidth}{!}{
    \includegraphics{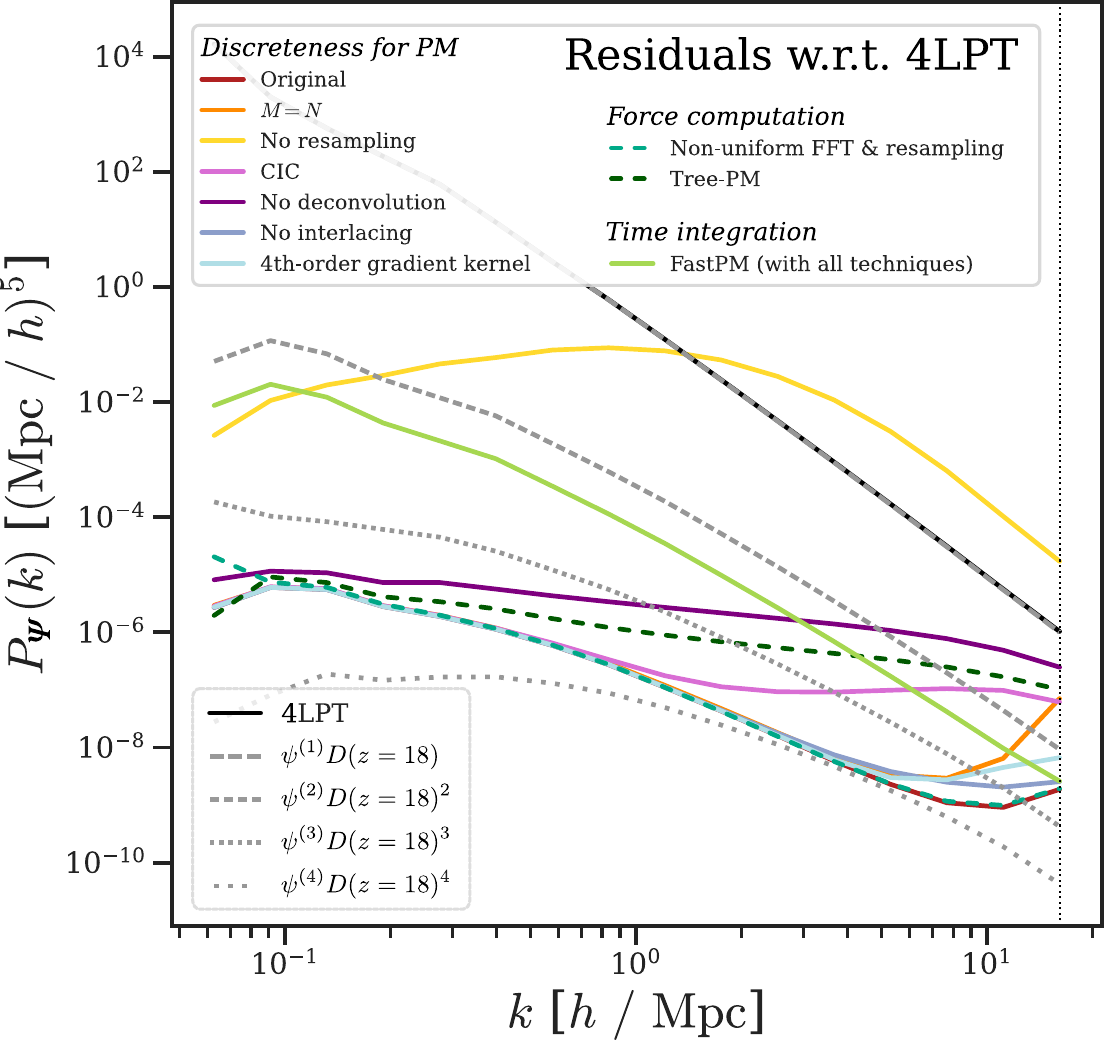}    
    }
    \caption{Power spectra of the displacement residuals between the \textsc{PowerFrog} results and 4LPT for the different cases considered in the ablation study shown in Figs.~\ref{fig:displacement_plot_ablation} and \ref{fig:displacement_plot_fastpm}, revealing which scales are affected by omitting each of the discreteness reduction techniques, by employing \textsc{FastPM} instead of \textsc{PowerFrog}, and by using force computation methods other than PM. The gray dashed/dotted lines show the displacement power spectra of the 1$-$4LPT contributions to the 4LPT displacement, and the black line indicates the total 4LPT displacement power spectrum (which can be seen to be dominated by the 1LPT term). The dotted vertical line shows the Nyquist mode for $N = 512^3$.}
    \label{fig:displacement_residual_rms}
\end{figure*}

\subsection{Displacement residual power spectra}

\noindent In order to quantitatively assess which scales are affected by the different discreteness reduction techniques, we plot the power spectrum of the residuals between the 1-step displacement and 4LPT for each case in Fig.~\ref{fig:displacement_residual_rms}. 
The residual displacement power spectrum in the case of omitting the resampling peaks at intermediate scales $k \approx 1 \, h \ / \ \text{Mpc}$ and lies above the 4LPT displacement power spectrum on smaller scales, explaining why the residuals towards the different LPT orders are visually indistinguishable in Fig.~\ref{fig:displacement_plot_ablation} in this case. Leaving out the deconvolution affects all scales, but most significantly small scales (in relative terms), where the resulting residual dominates over the 2LPT contribution, for which reason the residuals towards 2LPT and 3LPT look very similar in Fig.~\ref{fig:displacement_plot_ablation}. Using a fourth-order gradient kernel or a coarser PM grid ($M = N$) only adds power on the smallest scales near the Nyquist scale, however with a significant amplitude in the latter case which surpasses that of the 2LPT contribution. With a \textsc{FastPM} step, the shape of the power spectrum of the displacement residual w.r.t.\ 4LPT is extremely similar to the power spectrum of the 2LPT contribution, indicative of the fact that \textsc{FastPM} does not correctly capture the 2LPT term. 

\subsection{Curl generation due to discreteness and truncation}

\noindent The chosen initial conditions for our simulations imply initial potentiality of the velocity and thus $\bnabla_{\vecb{x}} \times \vecb{P} = 0$. Formulated in Lagrangian coordinates, this Eulerian constraint turns into the so-called Cauchy invariants \cite{Rampf:2012,Zheligovsky:2014}:
\begin{equation} \label{eqs:cauchy}
    {\cal C}_i := \varepsilon_{ijk} \, \partial_{q_j} X_{l} \, \partial_{q_k} P_{l} = 0,
\end{equation}
where $\varepsilon_{ijk}$ is the Levi--Civita symbol, and summation over repeated indices is implied.

The exact solution satisfies ${\cal C}_i = 0$ due to Helmholtz's third theorem, which states that an initially irrotational fluid will remain irrotational if only subject to conservative forces.
Within a perturbative setup at $n$th order in LPT, however, Eqs.\,\eqref{eqs:cauchy} are only satisfied to $n$th order, meaning that in general a truncation error remains. This has been analyzed in Refs.~\cite{Uhlemann:2018gzz,Michaux:2021,Rampf:2021} within various setups, where it was found that the $n$LPT truncation error of~\eqref{eqs:cauchy} is proportional to $D^n$ for $n>1$ (at sufficiently early times).
By contrast, for $n=1$ in LPT, there is no perturbative truncation error and thus ${\cal C}_i = 0$ up to machine precision.

While we leave a detailed study of the curl generation intrinsic to numerical time integrators for cosmological simulations for future work, it is expected that the discrete approximation of the particle trajectories predicted by \textsc{PowerFrog} (and other integrators) spuriously generate vorticity (similarly to $n$LPT for $n \geq 2$), even in the absence of discreteness, i.e.\ for $\varphi_N = \varphi$. In addition, as shown in Ref.~\cite{Marcos:2006}, particle discreteness leads to the spurious generation of curl, for which reason the conservation of ${\cal C}_i$ will be violated. In the following, we assess both effects.

\begin{figure}
    \centering
    \resizebox{1\columnwidth}{!}{
    \includegraphics{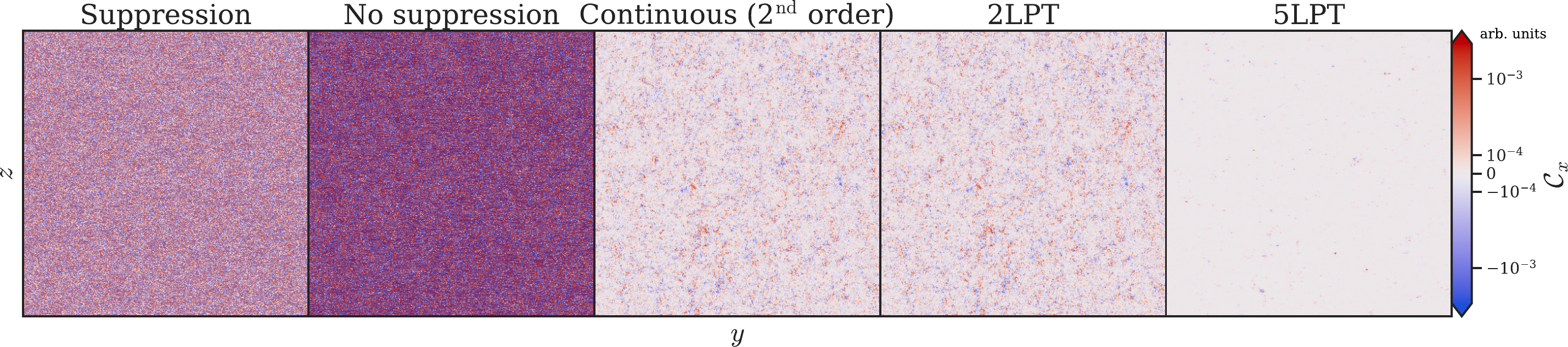}
    }
    \caption{A slice of the Cauchy invariant ${\cal C}_x$ after a single \textsc{PowerFrog} $N$-body step from $z = \infty$ to $18$ (with and without discreteness suppressed), and the same for 2LPT and 5LPT. The central panel shows ${\cal C}_x$ for a single \textsc{PowerFrog} step where the PM acceleration in the kick is replaced by a second-order approximation in terms of LPT terms (and hence without discreteness).}
    \label{fig:cauchy}
\end{figure}

Figure~\ref{fig:cauchy} shows a slice of the Cauchy invariant ${\cal C}_x$ at $z = 18$ in the $y-z$ coordinate plane. For 2LPT, a residual is visible, as expected. For 5LPT, which is much closer to the exact irrotational solution (as the LPT expansion is still valid in the single-stream regime at $z = 18$), the residual is very small. Interestingly, even with discreteness suppression, the Cauchy invariant ${\cal C}_x$ after a \textsc{PowerFrog} step is dominated by noise, and no characteristic signature inherent to the \textsc{PowerFrog} integrator is visible. This might be due to the fact that the computation of the Cauchy invariants involves two spatial derivatives, which amplifies the effect of small-scale noise. Therefore, in order to reveal the intrinsic \textsc{PowerFrog} signature, we also show the results with \textsc{PowerFrog} when replacing the discrete potential computed with PM in the kick by the second-order approximation w.r.t.\ $D$ of the potential at that time as provided by LPT (central panel). In that case, ${\cal C}_x$ is extremely similar to the 2LPT case. As can be seen in Figs.~\ref{fig:pk_plot_z_0}, \ref{fig:stats_z_0}, and \ref{fig:stats_z_3}, this spurious noise does not significantly affect summary statistics at late times.

\subsection{Ablation study for a step from \texorpdfstring{$\boldsymbol{z = \infty}$}{redshift infinity} to \texorpdfstring{$\boldsymbol{z = 36}$}{redshift 36}}
\begin{figure}
    \centering
    \resizebox{0.57\columnwidth}{!}{
    \includegraphics{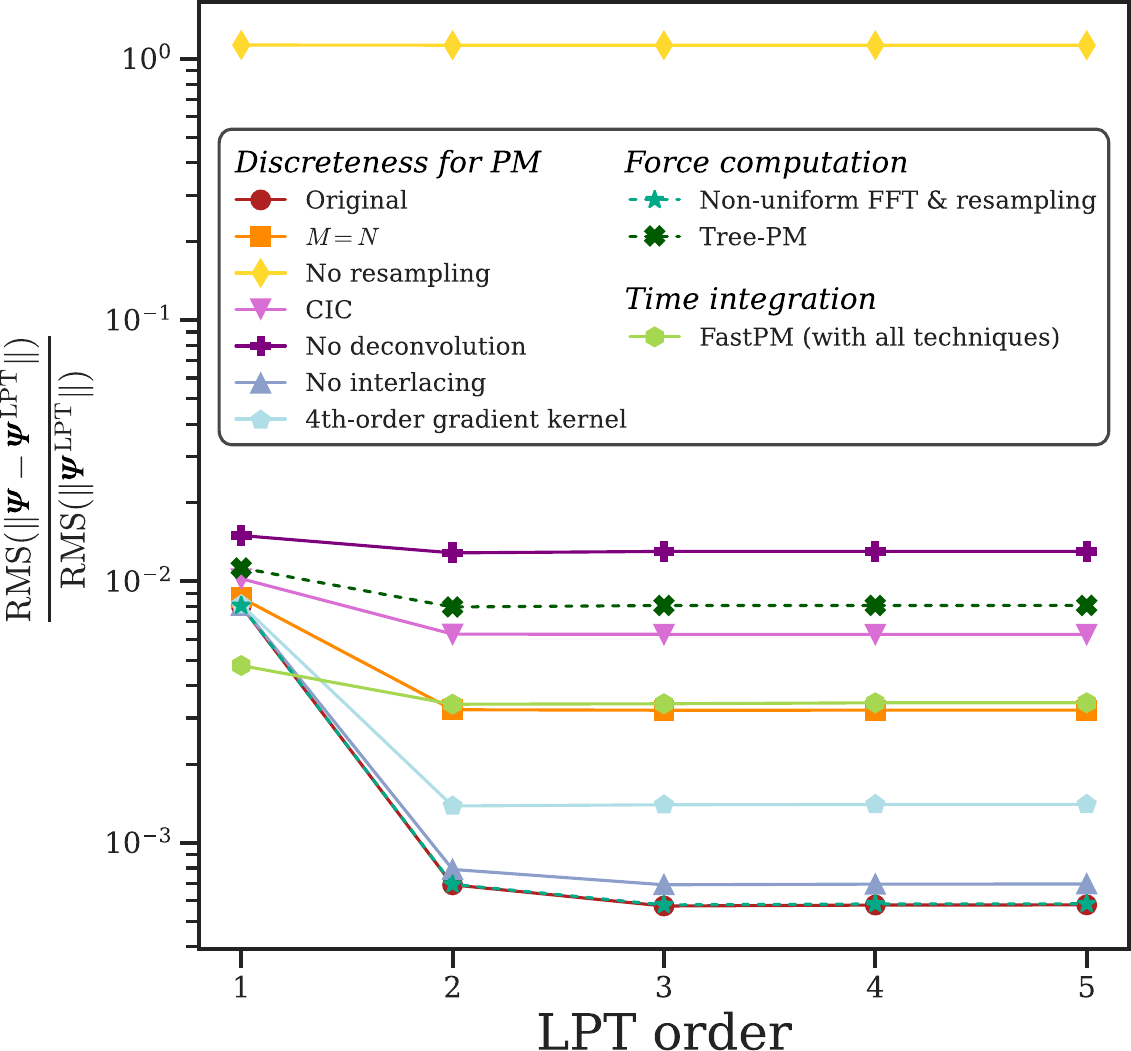}
    }
    \caption{Same as Fig.~\ref{fig:psi_rms_plot} in the main body, but for a single step from $z = \infty$ to $z = 36$ instead of $z = 18$: relative RMS error between the 1-step $N$-body simulation and different LPT orders when using the \textsc{PowerFrog} integrator and applying all discreteness reduction techniques (`Original'), when omitting one technique at a time, when performing a \textsc{FastPM} DKD step instead of a \textsc{PowerFrog} step (applying all discreteness reduction techniques), and when using the non-uniform FFT (with resampling) or Tree-PM (without resampling) instead of a local PM mass assignment kernel.}
    \label{fig:psi_rms_plot_z_36}
\end{figure}

\noindent Finally, we study how the impact of the different discreteness suppression techniques varies when changing the end time of the single $N$-body step. Figure~\ref{fig:psi_rms_plot_z_36} shows again the RMS errors when omitting each discreteness suppression technique at a time when performing a single time step from $z = \infty$, but now to $z = 36$ instead of $z = 18$ as shown in Fig.~\ref{fig:psi_rms_plot} in the main body. Going to this higher redshift slightly affects the order of importance of the different techniques: while the resampling followed by the deconvolution still have the largest effect, using CIC instead of PCS mass assignment now leads to a larger error, reflecting the increased need to suppress small-scale discreteness effects at early times. When applying all techniques, the relative error between a single \textsc{PowerFrog} step and 3LPT at $z = 36$ is $< 0.06\%$. Since the 3LPT contribution at $z = 36$ is still small, so is the difference between the 2LPT and 3LPT residuals for the `Original' case.

\section{Additional results at late times}
\subsection{Cross-spectrum and bispectrum}

\noindent In the main body, we used the particle positions and momenta computed with a single \textsc{PowerFrog} step from $z = \infty$ to $z = 36$ as initial conditions for a standard $N$-body simulation with the \textsc{Gadget-4} simulation code and compared the resulting power spectrum at $z = 0$ to its counterparts from LPT-initialized simulations. We found that even if no discreteness suppression techniques are employed for the initialization step, the power spectra agree to within $1\%$ on all scales (see Fig.~\ref{fig:pk_plot_z_0}). However, the power spectrum alone is not a sufficient statistic for determining if the quality of the $z = 0$ field computed with the \textsc{PowerFrog} initial conditions is satisfactory. 

\begin{figure}
    \centering
    \resizebox{0.66\columnwidth}{!}{
    \includegraphics{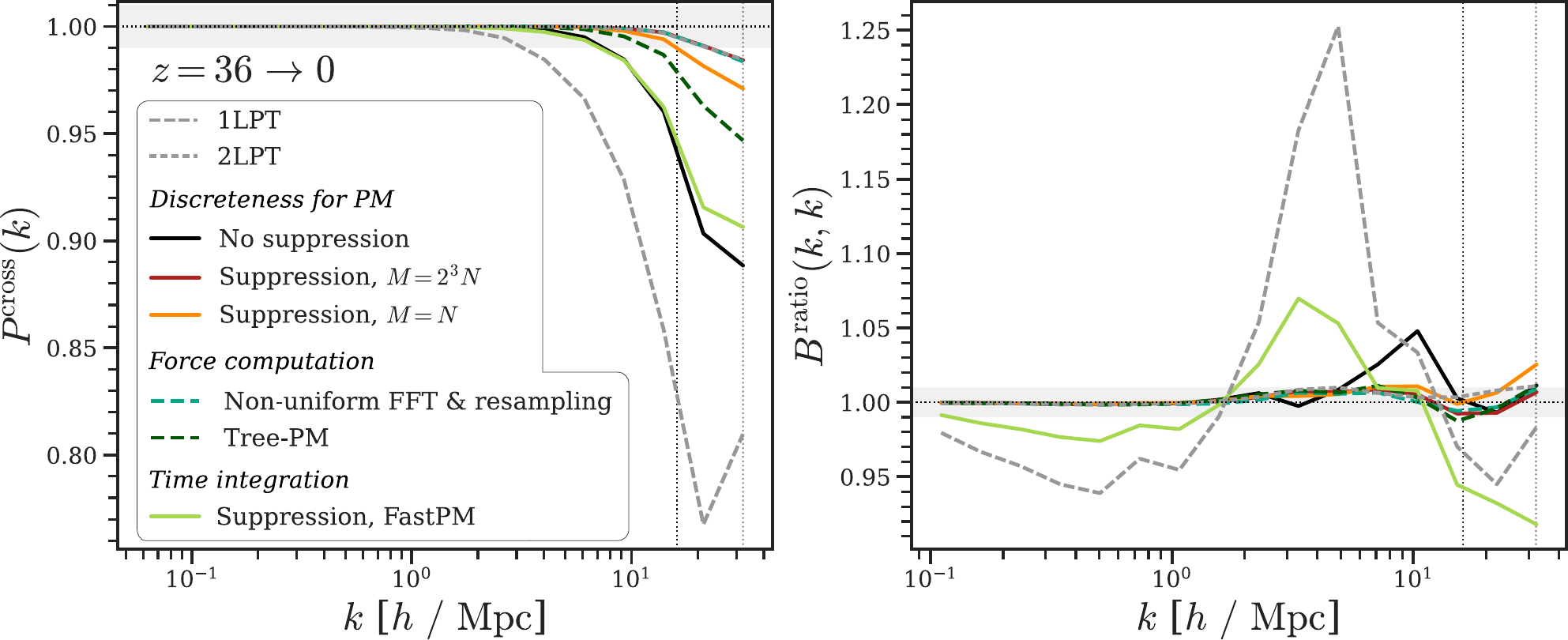}
    }
    \caption{Cross-power spectrum (\textit{left}) and equilateral bispectrum ratio (\textit{right}) between the $z = 0$ density fields with initial conditions generated at $z = 36$ (either with LPT or with a single $N$-body step from $z = \infty$) w.r.t.\ 3LPT initial conditions. The results for the power spectrum are shown in Fig.~\ref{fig:pk_plot_z_0} in the main part.}
    \label{fig:stats_z_0}    
\end{figure}

Figure~\ref{fig:stats_z_0} shows the (normalized) cross-spectrum for each simulation w.r.t.\ the 3LPT-initialized simulation, which we take as our reference. Unlike for the power spectrum, the impact of the discreteness suppression on the cross-spectrum is significant: omitting the discreteness suppression causes a drastic drop in cross-power on small scales, comparable in magnitude to the impact of using \textsc{FastPM} instead of \textsc{PowerFrog} (while keeping all discreteness suppression techniques). This implies that the coherence between the discrete and continuous phases is irretrievably corrupted by the discreteness. The cross-spectra for our discreteness-suppressed PM baseline with $M = 2^3 N$ grid cells and for the non-uniform FFT (with resampling) are virtually the same as for the 2LPT ICs, while the Tree-PM ICs lead to a slightly worse cross-spectrum (however, still with an error $< 1\%$ almost down to the particle Nyquist frequency).

As a further check, we plot the equilateral bispectra in the right panel of Fig.~\ref{fig:stats_z_0}. For all ICs except for 1LPT, \textsc{FastPM} (with discreteness suppression), and \textsc{PowerFrog} without discreteness suppression, the bispectrum errors are approximately within 1\% on all scales down to the particle Nyquist scale.

\subsection{Statistics at \texorpdfstring{$\boldsymbol{z = 3}$}{redshift 3}}

\begin{figure}
    \centering
    \resizebox{1\columnwidth}{!}{
    \includegraphics{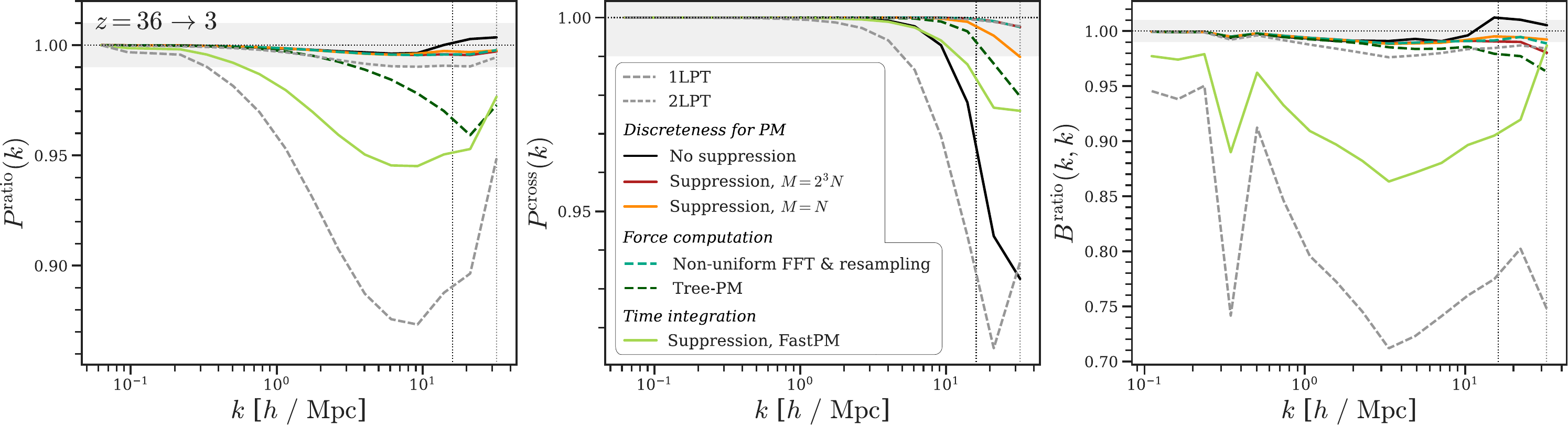}
    }
    \caption{Same statistics as in Figs.~\ref{fig:pk_plot_z_0} and \ref{fig:stats_z_0}, but evaluated at $z = 3$.}
    \label{fig:stats_z_3}    
\end{figure}

\noindent To study the time dependence of these statistics, we also show results at $z = 3$; see Fig.~\ref{fig:stats_z_3}. 
Also here, the agreement between the power spectra with 1-step \textsc{PowerFrog} initial conditions and with 3LPT is excellent---regardless of the discreteness suppression---and superior to the 2LPT initial conditions, which produce a slight suppression of power on small scales due to transients that have not fully decayed by $z = 3$. Similar to the $z = 0$ case, the equilateral bispectra match well; however, the cross-spectrum is strongly affected by the discreteness noise on small scales in the `No suppression' case. 

\begin{thebibliography}{69}%
\makeatletter
\providecommand \@ifxundefined [1]{%
 \@ifx{#1\undefined}
}%
\providecommand \@ifnum [1]{%
 \ifnum #1\expandafter \@firstoftwo
 \else \expandafter \@secondoftwo
 \fi
}%
\providecommand \@ifx [1]{%
 \ifx #1\expandafter \@firstoftwo
 \else \expandafter \@secondoftwo
 \fi
}%
\providecommand \natexlab [1]{#1}%
\providecommand \enquote  [1]{``#1''}%
\providecommand \bibnamefont  [1]{#1}%
\providecommand \bibfnamefont [1]{#1}%
\providecommand \citenamefont [1]{#1}%
\providecommand \href@noop [0]{\@secondoftwo}%
\providecommand \href [0]{\begingroup \@sanitize@url \@href}%
\providecommand \@href[1]{\@@startlink{#1}\@@href}%
\providecommand \@@href[1]{\endgroup#1\@@endlink}%
\providecommand \@sanitize@url [0]{\catcode `\\12\catcode `\$12\catcode
  `\&12\catcode `\#12\catcode `\^12\catcode `\_12\catcode `\%12\relax}%
\providecommand \@@startlink[1]{}%
\providecommand \@@endlink[0]{}%
\providecommand \url  [0]{\begingroup\@sanitize@url \@url }%
\providecommand \@url [1]{\endgroup\@href {#1}{\urlprefix }}%
\providecommand \urlprefix  [0]{URL }%
\providecommand \Eprint [0]{\href }%
\providecommand \doibase [0]{https://doi.org/}%
\providecommand \selectlanguage [0]{\@gobble}%
\providecommand \bibinfo  [0]{\@secondoftwo}%
\providecommand \bibfield  [0]{\@secondoftwo}%
\providecommand \translation [1]{[#1]}%
\providecommand \BibitemOpen [0]{}%
\providecommand \bibitemStop [0]{}%
\providecommand \bibitemNoStop [0]{.\EOS\space}%
\providecommand \EOS [0]{\spacefactor3000\relax}%
\providecommand \BibitemShut  [1]{\csname bibitem#1\endcsname}%
\let\auto@bib@innerbib\@empty
\bibitem [{\citenamefont {Peebles}(1980)}]{Peebles:1980}%
  \BibitemOpen
  \bibfield  {author} {\bibinfo {author} {\bibfnamefont {P.}~\bibnamefont
  {Peebles}},\ }\href@noop {} {\emph {\bibinfo {title} {The Large-scale
  Structure of the Universe}}},\ Princeton Series in Physics\ (\bibinfo
  {publisher} {Princeton University Press},\ \bibinfo {year}
  {1980})\BibitemShut {NoStop}%
\bibitem [{\citenamefont {Bernardeau}\ \emph {et~al.}(2002)\citenamefont
  {Bernardeau}, \citenamefont {Colombi}, \citenamefont {Gaztanaga},\ and\
  \citenamefont {Scoccimarro}}]{bernardeau2002large}%
  \BibitemOpen
  \bibfield  {author} {\bibinfo {author} {\bibfnamefont {F.}~\bibnamefont
  {Bernardeau}}, \bibinfo {author} {\bibfnamefont {S.}~\bibnamefont {Colombi}},
  \bibinfo {author} {\bibfnamefont {E.}~\bibnamefont {Gaztanaga}},\ and\
  \bibinfo {author} {\bibfnamefont {R.}~\bibnamefont {Scoccimarro}},\ }\href
  {https://doi.org/10.1016/S0370-1573(02)00135-7} {\bibfield  {journal}
  {\bibinfo  {journal} {Phys. Rep.}\ }\textbf {\bibinfo {volume} {367}},\
  \bibinfo {pages} {1} (\bibinfo {year} {2002})}\BibitemShut {NoStop}%
\bibitem [{\citenamefont {Rampf}(2021)}]{Rampf:2021rqu}%
  \BibitemOpen
  \bibfield  {author} {\bibinfo {author} {\bibfnamefont {C.}~\bibnamefont
  {Rampf}},\ }\href {https://doi.org/10.1007/s41614-021-00055-z} {\bibfield
  {journal} {\bibinfo  {journal} {Rev. Mod. Plasma Phys.}\ }\textbf {\bibinfo
  {volume} {5}},\ \bibinfo {pages} {10} (\bibinfo {year} {2021})},\ \Eprint
  {https://arxiv.org/abs/2110.06265} {arXiv:2110.06265 [astro-ph.CO]}
  \BibitemShut {NoStop}%
\bibitem [{\citenamefont {{Angulo}}\ and\ \citenamefont
  {{Hahn}}(2022)}]{2022LRCA....8....1A}%
  \BibitemOpen
  \bibfield  {author} {\bibinfo {author} {\bibfnamefont {R.~E.}\ \bibnamefont
  {{Angulo}}}\ and\ \bibinfo {author} {\bibfnamefont {O.}~\bibnamefont
  {{Hahn}}},\ }\href {https://doi.org/10.1007/s41115-021-00013-z} {\bibfield
  {journal} {\bibinfo  {journal} {Living rev. comput. astrophys.}\ }\textbf
  {\bibinfo {volume} {8}},\ \bibinfo {eid} {1} (\bibinfo {year} {2022})},\
  \Eprint {https://arxiv.org/abs/2112.05165} {arXiv:2112.05165 [astro-ph.CO]}
  \BibitemShut {NoStop}%
\bibitem [{\citenamefont {{Zel'dovich}}(1970)}]{Zeldovich:1970}%
  \BibitemOpen
  \bibfield  {author} {\bibinfo {author} {\bibfnamefont {{\relax Ya.
  B}.}~\bibnamefont {{Zel'dovich}}},\ }\href@noop {} {\bibfield  {journal}
  {\bibinfo  {journal} {\aap}\ }\textbf {\bibinfo {volume} {5}},\ \bibinfo
  {pages} {84} (\bibinfo {year} {1970})}\BibitemShut {NoStop}%
\bibitem [{\citenamefont {{Buchert}}\ and\ \citenamefont
  {{Ehlers}}(1993)}]{Buchert:1993}%
  \BibitemOpen
  \bibfield  {author} {\bibinfo {author} {\bibfnamefont {T.}~\bibnamefont
  {{Buchert}}}\ and\ \bibinfo {author} {\bibfnamefont {J.}~\bibnamefont
  {{Ehlers}}},\ }\href {https://doi.org/10.1093/mnras/264.2.375} {\bibfield
  {journal} {\bibinfo  {journal} {\mnras}\ }\textbf {\bibinfo {volume} {264}},\
  \bibinfo {pages} {375} (\bibinfo {year} {1993})}\BibitemShut {NoStop}%
\bibitem [{\citenamefont {{Bouchet}}\ \emph {et~al.}(1995)\citenamefont
  {{Bouchet}}, \citenamefont {{Colombi}}, \citenamefont {{Hivon}},\ and\
  \citenamefont {{Juszkiewicz}}}]{Bouchet:1995}%
  \BibitemOpen
  \bibfield  {author} {\bibinfo {author} {\bibfnamefont {F.~R.}\ \bibnamefont
  {{Bouchet}}}, \bibinfo {author} {\bibfnamefont {S.}~\bibnamefont
  {{Colombi}}}, \bibinfo {author} {\bibfnamefont {E.}~\bibnamefont {{Hivon}}},\
  and\ \bibinfo {author} {\bibfnamefont {R.}~\bibnamefont {{Juszkiewicz}}},\
  }\href@noop {} {\bibfield  {journal} {\bibinfo  {journal} {\aap}\ }\textbf
  {\bibinfo {volume} {296}},\ \bibinfo {pages} {575} (\bibinfo {year}
  {1995})},\ \Eprint {https://arxiv.org/abs/astro-ph/9406013}
  {arXiv:astro-ph/9406013 [astro-ph]} \BibitemShut {NoStop}%
\bibitem [{\citenamefont {{Ehlers}}\ and\ \citenamefont
  {{Buchert}}(1997)}]{1997GReGr..29..733E}%
  \BibitemOpen
  \bibfield  {author} {\bibinfo {author} {\bibfnamefont {J.}~\bibnamefont
  {{Ehlers}}}\ and\ \bibinfo {author} {\bibfnamefont {T.}~\bibnamefont
  {{Buchert}}},\ }\href {https://doi.org/10.1023/A:1018885922682} {\bibfield
  {journal} {\bibinfo  {journal} {Gen. Relativ. Gravit}\ }\textbf {\bibinfo
  {volume} {29}},\ \bibinfo {pages} {733} (\bibinfo {year} {1997})},\ \Eprint
  {https://arxiv.org/abs/astro-ph/9609036} {arXiv:astro-ph/9609036 [astro-ph]}
  \BibitemShut {NoStop}%
\bibitem [{\citenamefont {{Rampf}}(2012)}]{Rampf:2012}%
  \BibitemOpen
  \bibfield  {author} {\bibinfo {author} {\bibfnamefont {C.}~\bibnamefont
  {{Rampf}}},\ }\href {https://doi.org/10.1088/1475-7516/2012/12/004}
  {\bibfield  {journal} {\bibinfo  {journal} {\jcap}\ }\textbf {\bibinfo
  {volume} {2012}},\ \bibinfo {eid} {004} (\bibinfo {year} {2012})},\ \Eprint
  {https://arxiv.org/abs/1205.5274} {arXiv:1205.5274 [astro-ph.CO]}
  \BibitemShut {NoStop}%
\bibitem [{\citenamefont {{Zheligovsky}}\ and\ \citenamefont
  {{Frisch}}(2014)}]{Zheligovsky:2014}%
  \BibitemOpen
  \bibfield  {author} {\bibinfo {author} {\bibfnamefont {V.}~\bibnamefont
  {{Zheligovsky}}}\ and\ \bibinfo {author} {\bibfnamefont {U.}~\bibnamefont
  {{Frisch}}},\ }\href {https://doi.org/10.1017/jfm.2014.221} {\bibfield
  {journal} {\bibinfo  {journal} {J. Fluid Mech.}\ }\textbf {\bibinfo {volume}
  {749}},\ \bibinfo {pages} {404} (\bibinfo {year} {2014})},\ \Eprint
  {https://arxiv.org/abs/1312.6320} {arXiv:1312.6320 [math.AP]} \BibitemShut
  {NoStop}%
\bibitem [{\citenamefont {{Matsubara}}(2015)}]{Matsubara:2015}%
  \BibitemOpen
  \bibfield  {author} {\bibinfo {author} {\bibfnamefont {T.}~\bibnamefont
  {{Matsubara}}},\ }\href {https://doi.org/10.1103/PhysRevD.92.023534}
  {\bibfield  {journal} {\bibinfo  {journal} {\prd}\ }\textbf {\bibinfo
  {volume} {92}},\ \bibinfo {eid} {023534} (\bibinfo {year} {2015})},\ \Eprint
  {https://arxiv.org/abs/1505.01481} {arXiv:1505.01481 [astro-ph.CO]}
  \BibitemShut {NoStop}%
\bibitem [{\citenamefont {{Saga}}\ \emph {et~al.}(2018)\citenamefont {{Saga}},
  \citenamefont {{Taruya}},\ and\ \citenamefont
  {{Colombi}}}]{2018PhRvL.121x1302S}%
  \BibitemOpen
  \bibfield  {author} {\bibinfo {author} {\bibfnamefont {S.}~\bibnamefont
  {{Saga}}}, \bibinfo {author} {\bibfnamefont {A.}~\bibnamefont {{Taruya}}},\
  and\ \bibinfo {author} {\bibfnamefont {S.}~\bibnamefont {{Colombi}}},\ }\href
  {https://doi.org/10.1103/PhysRevLett.121.241302} {\bibfield  {journal}
  {\bibinfo  {journal} {\prl}\ }\textbf {\bibinfo {volume} {121}},\ \bibinfo
  {eid} {241302} (\bibinfo {year} {2018})},\ \Eprint
  {https://arxiv.org/abs/1805.08787} {arXiv:1805.08787 [astro-ph.CO]}
  \BibitemShut {NoStop}%
\bibitem [{\citenamefont {Colombi}(2015)}]{Colombi2015}%
  \BibitemOpen
  \bibfield  {author} {\bibinfo {author} {\bibfnamefont {S.}~\bibnamefont
  {Colombi}},\ }\href {https://doi.org/10.1093/mnras/stu2308} {\bibfield
  {journal} {\bibinfo  {journal} {\mnras}\ }\textbf {\bibinfo {volume} {446}},\
  \bibinfo {pages} {2902} (\bibinfo {year} {2015})},\ \Eprint
  {https://arxiv.org/abs/1411.4165} {arXiv:1411.4165 [astro-ph.CO]}
  \BibitemShut {NoStop}%
\bibitem [{\citenamefont {Taruya}\ and\ \citenamefont
  {Colombi}(2017)}]{Taruya2017}%
  \BibitemOpen
  \bibfield  {author} {\bibinfo {author} {\bibfnamefont {A.}~\bibnamefont
  {Taruya}}\ and\ \bibinfo {author} {\bibfnamefont {S.}~\bibnamefont
  {Colombi}},\ }\href {https://doi.org/10.1093/MNRAS/STX1501} {\bibfield
  {journal} {\bibinfo  {journal} {\mnras}\ }\textbf {\bibinfo {volume} {470}},\
  \bibinfo {pages} {4858} (\bibinfo {year} {2017})},\ \Eprint
  {https://arxiv.org/abs/1701.09088} {arXiv:1701.09088 [astro-ph.CO]}
  \BibitemShut {NoStop}%
\bibitem [{\citenamefont {{Rampf}}\ and\ \citenamefont
  {{Hahn}}(2021)}]{Rampf:2021}%
  \BibitemOpen
  \bibfield  {author} {\bibinfo {author} {\bibfnamefont {C.}~\bibnamefont
  {{Rampf}}}\ and\ \bibinfo {author} {\bibfnamefont {O.}~\bibnamefont
  {{Hahn}}},\ }\href {https://doi.org/10.1093/mnrasl/slaa198} {\bibfield
  {journal} {\bibinfo  {journal} {\mnras}\ }\textbf {\bibinfo {volume} {501}},\
  \bibinfo {pages} {L71} (\bibinfo {year} {2021})},\ \Eprint
  {https://arxiv.org/abs/2010.12584} {arXiv:2010.12584 [astro-ph.CO]}
  \BibitemShut {NoStop}%
\bibitem [{\citenamefont {{Saga}}\ \emph {et~al.}(2022)\citenamefont {{Saga}},
  \citenamefont {{Taruya}},\ and\ \citenamefont
  {{Colombi}}}]{2022A&A...664A...3S}%
  \BibitemOpen
  \bibfield  {author} {\bibinfo {author} {\bibfnamefont {S.}~\bibnamefont
  {{Saga}}}, \bibinfo {author} {\bibfnamefont {A.}~\bibnamefont {{Taruya}}},\
  and\ \bibinfo {author} {\bibfnamefont {S.}~\bibnamefont {{Colombi}}},\ }\href
  {https://doi.org/10.1051/0004-6361/202142756} {\bibfield  {journal} {\bibinfo
   {journal} {\aap}\ }\textbf {\bibinfo {volume} {664}},\ \bibinfo {eid} {A3}
  (\bibinfo {year} {2022})},\ \Eprint {https://arxiv.org/abs/2111.08836}
  {arXiv:2111.08836 [astro-ph.CO]} \BibitemShut {NoStop}%
\bibitem [{\citenamefont {Baumann}\ \emph {et~al.}(2012)\citenamefont
  {Baumann}, \citenamefont {Nicolis}, \citenamefont {Senatore},\ and\
  \citenamefont {Zaldarriaga}}]{baumann2012cosmological}%
  \BibitemOpen
  \bibfield  {author} {\bibinfo {author} {\bibfnamefont {D.}~\bibnamefont
  {Baumann}}, \bibinfo {author} {\bibfnamefont {A.}~\bibnamefont {Nicolis}},
  \bibinfo {author} {\bibfnamefont {L.}~\bibnamefont {Senatore}},\ and\
  \bibinfo {author} {\bibfnamefont {M.}~\bibnamefont {Zaldarriaga}},\ }\href
  {https://doi.org/10.1088/1475-7516/2012/07/051} {\bibfield  {journal}
  {\bibinfo  {journal} {\jcap}\ }\textbf {\bibinfo {volume} {07}},\ \bibinfo
  {pages} {051} (\bibinfo {year} {2012})},\ \Eprint
  {https://arxiv.org/abs/1004.2488} {arXiv:1004.2488 [astro-ph.CO]}
  \BibitemShut {NoStop}%
\bibitem [{\citenamefont {Carrasco}\ \emph {et~al.}(2012)\citenamefont
  {Carrasco}, \citenamefont {Hertzberg},\ and\ \citenamefont
  {Senatore}}]{carrasco2012effective}%
  \BibitemOpen
  \bibfield  {author} {\bibinfo {author} {\bibfnamefont {J.~J.~M.}\
  \bibnamefont {Carrasco}}, \bibinfo {author} {\bibfnamefont {M.~P.}\
  \bibnamefont {Hertzberg}},\ and\ \bibinfo {author} {\bibfnamefont
  {L.}~\bibnamefont {Senatore}},\ }\href
  {https://doi.org/10.1007/JHEP09(2012)082} {\bibfield  {journal} {\bibinfo
  {journal} {J. High Energy Phys.}\ }\textbf {\bibinfo {volume}
  {2012}}\bibfield  {number} {\bibinfo  {number} { (9)},\ \bibinfo {pages}
  {1}},\ }\Eprint {https://arxiv.org/abs/1206.2926} {arXiv:1206.2926
  [astro-ph.CO]} \BibitemShut {NoStop}%
\bibitem [{\citenamefont {{Cabass}}\ \emph {et~al.}(2023)\citenamefont
  {{Cabass}}, \citenamefont {{Ivanov}}, \citenamefont {{Lewandowski}},
  \citenamefont {{Mirbabayi}},\ and\ \citenamefont
  {{Simonovi{\'c}}}}]{cabass2023snowmass}%
  \BibitemOpen
  \bibfield  {author} {\bibinfo {author} {\bibfnamefont {G.}~\bibnamefont
  {{Cabass}}}, \bibinfo {author} {\bibfnamefont {M.~M.}\ \bibnamefont
  {{Ivanov}}}, \bibinfo {author} {\bibfnamefont {M.}~\bibnamefont
  {{Lewandowski}}}, \bibinfo {author} {\bibfnamefont {M.}~\bibnamefont
  {{Mirbabayi}}},\ and\ \bibinfo {author} {\bibfnamefont {M.}~\bibnamefont
  {{Simonovi{\'c}}}},\ }\href {https://doi.org/10.1016/j.dark.2023.101193}
  {\bibfield  {journal} {\bibinfo  {journal} {Phys. Dark Universe}\ }\textbf
  {\bibinfo {volume} {40}},\ \bibinfo {eid} {101193} (\bibinfo {year}
  {2023})},\ \Eprint {https://arxiv.org/abs/2203.08232} {arXiv:2203.08232
  [astro-ph.CO]} \BibitemShut {NoStop}%
\bibitem [{\citenamefont {{Joyce}}\ \emph {et~al.}(2005)\citenamefont
  {{Joyce}}, \citenamefont {{Marcos}}, \citenamefont {{Gabrielli}},
  \citenamefont {{Baertschiger}},\ and\ \citenamefont {{Sylos
  Labini}}}]{Joyce:2005}%
  \BibitemOpen
  \bibfield  {author} {\bibinfo {author} {\bibfnamefont {M.}~\bibnamefont
  {{Joyce}}}, \bibinfo {author} {\bibfnamefont {B.}~\bibnamefont {{Marcos}}},
  \bibinfo {author} {\bibfnamefont {A.}~\bibnamefont {{Gabrielli}}}, \bibinfo
  {author} {\bibfnamefont {T.}~\bibnamefont {{Baertschiger}}},\ and\ \bibinfo
  {author} {\bibfnamefont {F.}~\bibnamefont {{Sylos Labini}}},\ }\href
  {https://doi.org/10.1103/PhysRevLett.95.011304} {\bibfield  {journal}
  {\bibinfo  {journal} {\prl}\ }\textbf {\bibinfo {volume} {95}},\ \bibinfo
  {eid} {011304} (\bibinfo {year} {2005})},\ \Eprint
  {https://arxiv.org/abs/astro-ph/0504213} {arXiv:astro-ph/0504213 [astro-ph]}
  \BibitemShut {NoStop}%
\bibitem [{\citenamefont {{Marcos}}\ \emph {et~al.}(2006)\citenamefont
  {{Marcos}}, \citenamefont {{Baertschiger}}, \citenamefont {{Joyce}},
  \citenamefont {{Gabrielli}},\ and\ \citenamefont {{Sylos
  Labini}}}]{Marcos:2006}%
  \BibitemOpen
  \bibfield  {author} {\bibinfo {author} {\bibfnamefont {B.}~\bibnamefont
  {{Marcos}}}, \bibinfo {author} {\bibfnamefont {T.}~\bibnamefont
  {{Baertschiger}}}, \bibinfo {author} {\bibfnamefont {M.}~\bibnamefont
  {{Joyce}}}, \bibinfo {author} {\bibfnamefont {A.}~\bibnamefont
  {{Gabrielli}}},\ and\ \bibinfo {author} {\bibfnamefont {F.}~\bibnamefont
  {{Sylos Labini}}},\ }\href {https://doi.org/10.1103/PhysRevD.73.103507}
  {\bibfield  {journal} {\bibinfo  {journal} {\prd}\ }\textbf {\bibinfo
  {volume} {73}},\ \bibinfo {eid} {103507} (\bibinfo {year} {2006})},\ \Eprint
  {https://arxiv.org/abs/astro-ph/0601479} {arXiv:astro-ph/0601479 [astro-ph]}
  \BibitemShut {NoStop}%
\bibitem [{\citenamefont {{Garrison}}\ \emph {et~al.}(2016)\citenamefont
  {{Garrison}}, \citenamefont {{Eisenstein}}, \citenamefont {{Ferrer}},
  \citenamefont {{Metchnik}},\ and\ \citenamefont {{Pinto}}}]{Garrison:2016}%
  \BibitemOpen
  \bibfield  {author} {\bibinfo {author} {\bibfnamefont {L.~H.}\ \bibnamefont
  {{Garrison}}}, \bibinfo {author} {\bibfnamefont {D.~J.}\ \bibnamefont
  {{Eisenstein}}}, \bibinfo {author} {\bibfnamefont {D.}~\bibnamefont
  {{Ferrer}}}, \bibinfo {author} {\bibfnamefont {M.~V.}\ \bibnamefont
  {{Metchnik}}},\ and\ \bibinfo {author} {\bibfnamefont {P.~A.}\ \bibnamefont
  {{Pinto}}},\ }\href {https://doi.org/10.1093/mnras/stw1594} {\bibfield
  {journal} {\bibinfo  {journal} {\mnras}\ }\textbf {\bibinfo {volume} {461}},\
  \bibinfo {pages} {4125} (\bibinfo {year} {2016})},\ \Eprint
  {https://arxiv.org/abs/1605.02333} {arXiv:1605.02333 [astro-ph.CO]}
  \BibitemShut {NoStop}%
\bibitem [{\citenamefont {{Bouchet}}\ \emph {et~al.}(1992)\citenamefont
  {{Bouchet}}, \citenamefont {{Juszkiewicz}}, \citenamefont {{Colombi}},\ and\
  \citenamefont {{Pellat}}}]{1992ApJ...394L...5B}%
  \BibitemOpen
  \bibfield  {author} {\bibinfo {author} {\bibfnamefont {F.~R.}\ \bibnamefont
  {{Bouchet}}}, \bibinfo {author} {\bibfnamefont {R.}~\bibnamefont
  {{Juszkiewicz}}}, \bibinfo {author} {\bibfnamefont {S.}~\bibnamefont
  {{Colombi}}},\ and\ \bibinfo {author} {\bibfnamefont {R.}~\bibnamefont
  {{Pellat}}},\ }\href {https://doi.org/10.1086/186459} {\bibfield  {journal}
  {\bibinfo  {journal} {Astrophys. J. Lett.}\ }\textbf {\bibinfo {volume}
  {394}},\ \bibinfo {pages} {L5} (\bibinfo {year} {1992})}\BibitemShut
  {NoStop}%
\bibitem [{\citenamefont {{Michaux}}\ \emph {et~al.}(2021)\citenamefont
  {{Michaux}}, \citenamefont {{Hahn}}, \citenamefont {{Rampf}},\ and\
  \citenamefont {{Angulo}}}]{Michaux:2021}%
  \BibitemOpen
  \bibfield  {author} {\bibinfo {author} {\bibfnamefont {M.}~\bibnamefont
  {{Michaux}}}, \bibinfo {author} {\bibfnamefont {O.}~\bibnamefont {{Hahn}}},
  \bibinfo {author} {\bibfnamefont {C.}~\bibnamefont {{Rampf}}},\ and\ \bibinfo
  {author} {\bibfnamefont {R.~E.}\ \bibnamefont {{Angulo}}},\ }\href
  {https://doi.org/10.1093/mnras/staa3149} {\bibfield  {journal} {\bibinfo
  {journal} {\mnras}\ }\textbf {\bibinfo {volume} {500}},\ \bibinfo {pages}
  {663} (\bibinfo {year} {2021})},\ \Eprint {https://arxiv.org/abs/2008.09588}
  {arXiv:2008.09588 [astro-ph.CO]} \BibitemShut {NoStop}%
\bibitem [{\citenamefont {{Buchert}}(1994)}]{1994MNRAS.267..811B}%
  \BibitemOpen
  \bibfield  {author} {\bibinfo {author} {\bibfnamefont {T.}~\bibnamefont
  {{Buchert}}},\ }\href {https://doi.org/10.1093/mnras/267.4.811} {\bibfield
  {journal} {\bibinfo  {journal} {\mnras}\ }\textbf {\bibinfo {volume} {267}},\
  \bibinfo {pages} {811} (\bibinfo {year} {1994})},\ \Eprint
  {https://arxiv.org/abs/astro-ph/9309055} {arXiv:astro-ph/9309055 [astro-ph]}
  \BibitemShut {NoStop}%
\bibitem [{\citenamefont {{Catelan}}(1995)}]{1995MNRAS.276..115C}%
  \BibitemOpen
  \bibfield  {author} {\bibinfo {author} {\bibfnamefont {P.}~\bibnamefont
  {{Catelan}}},\ }\href {https://doi.org/10.1093/mnras/276.1.115} {\bibfield
  {journal} {\bibinfo  {journal} {\mnras}\ }\textbf {\bibinfo {volume} {276}},\
  \bibinfo {pages} {115} (\bibinfo {year} {1995})},\ \Eprint
  {https://arxiv.org/abs/astro-ph/9406016} {arXiv:astro-ph/9406016 [astro-ph]}
  \BibitemShut {NoStop}%
\bibitem [{\citenamefont {List}\ and\ \citenamefont
  {Hahn}(2023)}]{list2023perturbation}%
  \BibitemOpen
  \bibfield  {author} {\bibinfo {author} {\bibfnamefont {F.}~\bibnamefont
  {List}}\ and\ \bibinfo {author} {\bibfnamefont {O.}~\bibnamefont {Hahn}},\
  }\href@noop {} {\bibfield  {journal} {\bibinfo  {journal} {\preprint
  (arXiv:2301.09655)}\ } (\bibinfo {year} {2023})},\ \Eprint
  {https://arxiv.org/abs/2301.09655} {arXiv:2301.09655 [astro-ph.CO]}
  \BibitemShut {NoStop}%
\bibitem [{\citenamefont {Hockney}\ and\ \citenamefont
  {Eastwood}(1988)}]{HockneyEastwood}%
  \BibitemOpen
  \bibfield  {author} {\bibinfo {author} {\bibfnamefont {R.~W.}\ \bibnamefont
  {Hockney}}\ and\ \bibinfo {author} {\bibfnamefont {J.~W.}\ \bibnamefont
  {Eastwood}},\ }\href {https://doi.org/10.1201/9780367806934} {\emph {\bibinfo
  {title} {Computer Simulation Using Particles (1st ed.)}}}\ (\bibinfo
  {publisher} {CRC Press},\ \bibinfo {year} {1988})\BibitemShut {NoStop}%
\bibitem [{\citenamefont {Rampf}\ \emph {et~al.}(2022)\citenamefont {Rampf},
  \citenamefont {Schobesberger},\ and\ \citenamefont {Hahn}}]{Rampf:2022}%
  \BibitemOpen
  \bibfield  {author} {\bibinfo {author} {\bibfnamefont {C.}~\bibnamefont
  {Rampf}}, \bibinfo {author} {\bibfnamefont {S.~O.}\ \bibnamefont
  {Schobesberger}},\ and\ \bibinfo {author} {\bibfnamefont {O.}~\bibnamefont
  {Hahn}},\ }\href {https://doi.org/10.1093/mnras/stac2406} {\bibfield
  {journal} {\bibinfo  {journal} {\mnras}\ }\textbf {\bibinfo {volume} {516}},\
  \bibinfo {pages} {2840} (\bibinfo {year} {2022})},\ \Eprint
  {https://arxiv.org/abs/2205.11347} {arXiv:2205.11347 [astro-ph.CO]}
  \BibitemShut {NoStop}%
\bibitem [{\citenamefont {{Fasiello}}\ \emph
  {et~al.}(2022{\natexlab{a}})\citenamefont {{Fasiello}}, \citenamefont
  {{Fujita}},\ and\ \citenamefont {{Vlah}}}]{2022PhRvD.106l3504F}%
  \BibitemOpen
  \bibfield  {author} {\bibinfo {author} {\bibfnamefont {M.}~\bibnamefont
  {{Fasiello}}}, \bibinfo {author} {\bibfnamefont {T.}~\bibnamefont
  {{Fujita}}},\ and\ \bibinfo {author} {\bibfnamefont {Z.}~\bibnamefont
  {{Vlah}}},\ }\href {https://doi.org/10.1103/PhysRevD.106.123504} {\bibfield
  {journal} {\bibinfo  {journal} {\prd}\ }\textbf {\bibinfo {volume} {106}},\
  \bibinfo {eid} {123504} (\bibinfo {year} {2022}{\natexlab{a}})},\ \Eprint
  {https://arxiv.org/abs/2205.10026} {arXiv:2205.10026 [astro-ph.CO]}
  \BibitemShut {NoStop}%
\bibitem [{\citenamefont {{Feng}}\ \emph {et~al.}(2016)\citenamefont {{Feng}},
  \citenamefont {{Chu}}, \citenamefont {{Seljak}},\ and\ \citenamefont
  {{McDonald}}}]{Feng:2016}%
  \BibitemOpen
  \bibfield  {author} {\bibinfo {author} {\bibfnamefont {Y.}~\bibnamefont
  {{Feng}}}, \bibinfo {author} {\bibfnamefont {M.-Y.}\ \bibnamefont {{Chu}}},
  \bibinfo {author} {\bibfnamefont {U.}~\bibnamefont {{Seljak}}},\ and\
  \bibinfo {author} {\bibfnamefont {P.}~\bibnamefont {{McDonald}}},\ }\href
  {https://doi.org/10.1093/mnras/stw2123} {\bibfield  {journal} {\bibinfo
  {journal} {\mnras}\ }\textbf {\bibinfo {volume} {463}},\ \bibinfo {pages}
  {2273} (\bibinfo {year} {2016})},\ \Eprint {https://arxiv.org/abs/1603.00476}
  {arXiv:1603.00476 [astro-ph.CO]} \BibitemShut {NoStop}%
\bibitem [{\citenamefont {Brenier}\ \emph {et~al.}(2003)\citenamefont
  {Brenier}, \citenamefont {Frisch}, \citenamefont {H{\'{e}}non}, \citenamefont
  {Loeper}, \citenamefont {Matarrese}, \citenamefont {Mohayaee},\ and\
  \citenamefont {Sobolevskiǐ}}]{Brenier2003}%
  \BibitemOpen
  \bibfield  {author} {\bibinfo {author} {\bibfnamefont {Y.}~\bibnamefont
  {Brenier}}, \bibinfo {author} {\bibfnamefont {U.}~\bibnamefont {Frisch}},
  \bibinfo {author} {\bibfnamefont {M.}~\bibnamefont {H{\'{e}}non}}, \bibinfo
  {author} {\bibfnamefont {G.}~\bibnamefont {Loeper}}, \bibinfo {author}
  {\bibfnamefont {S.}~\bibnamefont {Matarrese}}, \bibinfo {author}
  {\bibfnamefont {R.}~\bibnamefont {Mohayaee}},\ and\ \bibinfo {author}
  {\bibfnamefont {A.}~\bibnamefont {Sobolevskiǐ}},\ }\href
  {https://doi.org/10.1046/j.1365-2966.2003.07106.x} {\bibfield  {journal}
  {\bibinfo  {journal} {\mnras}\ }\textbf {\bibinfo {volume} {346}},\ \bibinfo
  {pages} {501} (\bibinfo {year} {2003})},\ \Eprint
  {https://arxiv.org/abs/0304214} {arXiv:0304214 [astro-ph]} \BibitemShut
  {NoStop}%
\bibitem [{\citenamefont {{Frisch}}\ \emph {et~al.}(2002)\citenamefont
  {{Frisch}}, \citenamefont {{Matarrese}}, \citenamefont {{Mohayaee}},\ and\
  \citenamefont {{Sobolevski}}}]{2002Natur.417..260F}%
  \BibitemOpen
  \bibfield  {author} {\bibinfo {author} {\bibfnamefont {U.}~\bibnamefont
  {{Frisch}}}, \bibinfo {author} {\bibfnamefont {S.}~\bibnamefont
  {{Matarrese}}}, \bibinfo {author} {\bibfnamefont {R.}~\bibnamefont
  {{Mohayaee}}},\ and\ \bibinfo {author} {\bibfnamefont {A.}~\bibnamefont
  {{Sobolevski}}},\ }\href {https://doi.org/10.1038/417260a} {\bibfield
  {journal} {\bibinfo  {journal} {\nat}\ }\textbf {\bibinfo {volume} {417}},\
  \bibinfo {pages} {260} (\bibinfo {year} {2002})},\ \Eprint
  {https://arxiv.org/abs/astro-ph/0109483} {arXiv:astro-ph/0109483 [astro-ph]}
  \BibitemShut {NoStop}%
\bibitem [{\citenamefont {Rampf}(2019)}]{Rampf2019QuasiSpherical}%
  \BibitemOpen
  \bibfield  {author} {\bibinfo {author} {\bibfnamefont {C.}~\bibnamefont
  {Rampf}},\ }\href {https://doi.org/10.1093/mnras/stz372} {\bibfield
  {journal} {\bibinfo  {journal} {\mnras}\ }\textbf {\bibinfo {volume} {484}},\
  \bibinfo {pages} {5223} (\bibinfo {year} {2019})},\ \Eprint
  {https://arxiv.org/abs/1712.01878} {arXiv:1712.01878 [astro-ph.CO]}
  \BibitemShut {NoStop}%
\bibitem [{\citenamefont {Arnold}(1989)}]{Arnold:1989}%
  \BibitemOpen
  \bibfield  {author} {\bibinfo {author} {\bibfnamefont {V.}~\bibnamefont
  {Arnold}},\ }\href@noop {} {\emph {\bibinfo {title} {Mathematical methods of
  classical mechanics}}},\ Vol.~\bibinfo {volume} {60}\ (\bibinfo  {publisher}
  {Springer},\ \bibinfo {year} {1989})\BibitemShut {NoStop}%
\bibitem [{\citenamefont {Bravetti}\ \emph {et~al.}(2017)\citenamefont
  {Bravetti}, \citenamefont {Cruz},\ and\ \citenamefont
  {Tapias}}]{Bravetti2017a}%
  \BibitemOpen
  \bibfield  {author} {\bibinfo {author} {\bibfnamefont {A.}~\bibnamefont
  {Bravetti}}, \bibinfo {author} {\bibfnamefont {H.}~\bibnamefont {Cruz}},\
  and\ \bibinfo {author} {\bibfnamefont {D.}~\bibnamefont {Tapias}},\ }\href
  {https://doi.org/10.1016/j.aop.2016.11.003} {\bibfield  {journal} {\bibinfo
  {journal} {Ann. Phys.}\ }\textbf {\bibinfo {volume} {376}},\ \bibinfo {pages}
  {17} (\bibinfo {year} {2017})},\ \Eprint {https://arxiv.org/abs/1604.08266}
  {arXiv:1604.08266 [math-ph]} \BibitemShut {NoStop}%
\bibitem [{\citenamefont {Rampf}\ \emph {et~al.}(2021)\citenamefont {Rampf},
  \citenamefont {Uhlemann},\ and\ \citenamefont {Hahn}}]{Rampf2021TwoFluids}%
  \BibitemOpen
  \bibfield  {author} {\bibinfo {author} {\bibfnamefont {C.}~\bibnamefont
  {Rampf}}, \bibinfo {author} {\bibfnamefont {C.}~\bibnamefont {Uhlemann}},\
  and\ \bibinfo {author} {\bibfnamefont {O.}~\bibnamefont {Hahn}},\ }\href
  {https://doi.org/10.1093/mnras/staa3605} {\bibfield  {journal} {\bibinfo
  {journal} {\mnras}\ }\textbf {\bibinfo {volume} {503}},\ \bibinfo {pages}
  {406} (\bibinfo {year} {2021})},\ \Eprint {https://arxiv.org/abs/2008.09123}
  {arXiv:2008.09123 [astro-ph.CO]} \BibitemShut {NoStop}%
\bibitem [{\citenamefont {{Colombi}}(2021)}]{2021A&A...647A..66C}%
  \BibitemOpen
  \bibfield  {author} {\bibinfo {author} {\bibfnamefont {S.}~\bibnamefont
  {{Colombi}}},\ }\href {https://doi.org/10.1051/0004-6361/202039719}
  {\bibfield  {journal} {\bibinfo  {journal} {\aap}\ }\textbf {\bibinfo
  {volume} {647}},\ \bibinfo {eid} {A66} (\bibinfo {year} {2021})},\ \Eprint
  {https://arxiv.org/abs/2012.04409} {arXiv:2012.04409 [astro-ph.CO]}
  \BibitemShut {NoStop}%
\bibitem [{\citenamefont {{Feistl}}\ and\ \citenamefont
  {{Pickl}}(2023)}]{2023arXiv230706146F}%
  \BibitemOpen
  \bibfield  {author} {\bibinfo {author} {\bibfnamefont {M.}~\bibnamefont
  {{Feistl}}}\ and\ \bibinfo {author} {\bibfnamefont {P.}~\bibnamefont
  {{Pickl}}},\ }\href@noop {} {\bibfield  {journal} {\bibinfo  {journal}
  {\preprint (arXiv:2307.06146)}\ } (\bibinfo {year} {2023})},\ \Eprint
  {https://arxiv.org/abs/2307.06146} {arXiv:2307.06146 [math-ph]} \BibitemShut
  {NoStop}%
\bibitem [{\citenamefont {Bagla}(2002)}]{bagla2002treepm}%
  \BibitemOpen
  \bibfield  {author} {\bibinfo {author} {\bibfnamefont {J.~S.}\ \bibnamefont
  {Bagla}},\ }\href {https://doi.org/10.1007/BF02702282} {\bibfield  {journal}
  {\bibinfo  {journal} {J. Astrophys. Astron.}\ }\textbf {\bibinfo {volume}
  {23}},\ \bibinfo {pages} {185} (\bibinfo {year} {2002})},\ \Eprint
  {https://arxiv.org/abs/astro-ph/9911025} {arXiv:astro-ph/9911025}
  \BibitemShut {NoStop}%
\bibitem [{\citenamefont {Bode}\ and\ \citenamefont
  {Ostriker}(2003)}]{Bode2003}%
  \BibitemOpen
  \bibfield  {author} {\bibinfo {author} {\bibfnamefont {P.}~\bibnamefont
  {Bode}}\ and\ \bibinfo {author} {\bibfnamefont {J.~P.}\ \bibnamefont
  {Ostriker}},\ }\href {https://doi.org/10.1086/345538} {\bibfield  {journal}
  {\bibinfo  {journal} {\apjs}\ }\textbf {\bibinfo {volume} {145}},\ \bibinfo
  {pages} {1} (\bibinfo {year} {2003})},\ \Eprint
  {https://arxiv.org/abs/0302065} {arXiv:0302065 [astro-ph]} \BibitemShut
  {NoStop}%
\bibitem [{\citenamefont {{Hahn}}\ and\ \citenamefont
  {{Angulo}}(2016)}]{Hahn:2016}%
  \BibitemOpen
  \bibfield  {author} {\bibinfo {author} {\bibfnamefont {O.}~\bibnamefont
  {{Hahn}}}\ and\ \bibinfo {author} {\bibfnamefont {R.~E.}\ \bibnamefont
  {{Angulo}}},\ }\href {https://doi.org/10.1093/mnras/stv2304} {\bibfield
  {journal} {\bibinfo  {journal} {\mnras}\ }\textbf {\bibinfo {volume} {455}},\
  \bibinfo {pages} {1115} (\bibinfo {year} {2016})},\ \Eprint
  {https://arxiv.org/abs/1501.01959} {arXiv:1501.01959 [astro-ph.CO]}
  \BibitemShut {NoStop}%
\bibitem [{\citenamefont {St{\"u}cker}\ \emph {et~al.}(2018)\citenamefont
  {St{\"u}cker}, \citenamefont {Busch},\ and\ \citenamefont
  {White}}]{stucker2018median}%
  \BibitemOpen
  \bibfield  {author} {\bibinfo {author} {\bibfnamefont {J.}~\bibnamefont
  {St{\"u}cker}}, \bibinfo {author} {\bibfnamefont {P.}~\bibnamefont {Busch}},\
  and\ \bibinfo {author} {\bibfnamefont {S.~D.}\ \bibnamefont {White}},\ }\href
  {https://doi.org/10.1093/mnras/sty815} {\bibfield  {journal} {\bibinfo
  {journal} {\mnras}\ }\textbf {\bibinfo {volume} {477}},\ \bibinfo {pages}
  {3230} (\bibinfo {year} {2018})},\ \Eprint {https://arxiv.org/abs/1710.09881}
  {arXiv:1710.09881 [astro-ph.CO]} \BibitemShut {NoStop}%
\bibitem [{\citenamefont {Chaniotis}\ and\ \citenamefont
  {Poulikakos}(2004)}]{Chaniotis:2004}%
  \BibitemOpen
  \bibfield  {author} {\bibinfo {author} {\bibfnamefont {A.}~\bibnamefont
  {Chaniotis}}\ and\ \bibinfo {author} {\bibfnamefont {D.}~\bibnamefont
  {Poulikakos}},\ }\href
  {https://doi.org/https://doi.org/10.1016/j.jcp.2003.11.026} {\bibfield
  {journal} {\bibinfo  {journal} {\jcp}\ }\textbf {\bibinfo {volume} {197}},\
  \bibinfo {pages} {253} (\bibinfo {year} {2004})}\BibitemShut {NoStop}%
\bibitem [{\citenamefont {St{\"u}cker}\ \emph {et~al.}(2020)\citenamefont
  {St{\"u}cker}, \citenamefont {Hahn}, \citenamefont {Angulo},\ and\
  \citenamefont {White}}]{stucker2020simulating}%
  \BibitemOpen
  \bibfield  {author} {\bibinfo {author} {\bibfnamefont {J.}~\bibnamefont
  {St{\"u}cker}}, \bibinfo {author} {\bibfnamefont {O.}~\bibnamefont {Hahn}},
  \bibinfo {author} {\bibfnamefont {R.~E.}\ \bibnamefont {Angulo}},\ and\
  \bibinfo {author} {\bibfnamefont {S.~D.}\ \bibnamefont {White}},\ }\href
  {https://doi.org/10.1093/mnras/staa1468} {\bibfield  {journal} {\bibinfo
  {journal} {\mnras}\ }\textbf {\bibinfo {volume} {495}},\ \bibinfo {pages}
  {4943} (\bibinfo {year} {2020})},\ \Eprint {https://arxiv.org/abs/1909.00008}
  {arXiv:1909.00008 [astro-ph.CO]} \BibitemShut {NoStop}%
\bibitem [{jax(2021)}]{jaxfinufft}%
  \BibitemOpen
  \href@noop {} {\bibinfo {title} {jax-finufft: {J}{A}{X} bindings to the
  {F}latiron {I}nstitute {N}on-uniform {F}ast {F}ourier {T}ransform
  ({F}{I}{N}{U}{F}{F}{T}) library}},\ \bibinfo {howpublished}
  {\url{https://github.com/flatironinstitute/jax-finufft}} (\bibinfo {year}
  {2021})\BibitemShut {NoStop}%
\bibitem [{\citenamefont {Barnett}\ \emph {et~al.}(2019)\citenamefont
  {Barnett}, \citenamefont {Magland},\ and\ \citenamefont
  {af~Klinteberg}}]{barnett2019parallel}%
  \BibitemOpen
  \bibfield  {author} {\bibinfo {author} {\bibfnamefont {A.~H.}\ \bibnamefont
  {Barnett}}, \bibinfo {author} {\bibfnamefont {J.}~\bibnamefont {Magland}},\
  and\ \bibinfo {author} {\bibfnamefont {L.}~\bibnamefont {af~Klinteberg}},\
  }\href {https://doi.org/10.1137/18M120885X} {\bibfield  {journal} {\bibinfo
  {journal} {SIAM J. Sci. Comput.}\ }\textbf {\bibinfo {volume} {41}},\
  \bibinfo {pages} {C479} (\bibinfo {year} {2019})},\ \Eprint
  {https://arxiv.org/abs/1808.06736} {arXiv:1808.06736} \BibitemShut {NoStop}%
\bibitem [{\citenamefont {Barnett}(2021)}]{barnett2021aliasing}%
  \BibitemOpen
  \bibfield  {author} {\bibinfo {author} {\bibfnamefont {A.~H.}\ \bibnamefont
  {Barnett}},\ }\href {https://doi.org/10.1016/j.acha.2020.10.002} {\bibfield
  {journal} {\bibinfo  {journal} {Appl. Comput. Harmon. Anal.}\ }\textbf
  {\bibinfo {volume} {51}},\ \bibinfo {pages} {1} (\bibinfo {year} {2021})},\
  \Eprint {https://arxiv.org/abs/2001.09405} {arXiv:2001.09405} \BibitemShut
  {NoStop}%
\bibitem [{\citenamefont {Shih}\ \emph {et~al.}(2021)\citenamefont {Shih},
  \citenamefont {Wright}, \citenamefont {And{\'e}n}, \citenamefont {Blaschke},\
  and\ \citenamefont {Barnett}}]{shih2021cufinufft}%
  \BibitemOpen
  \bibfield  {author} {\bibinfo {author} {\bibfnamefont {Y.-h.}\ \bibnamefont
  {Shih}}, \bibinfo {author} {\bibfnamefont {G.}~\bibnamefont {Wright}},
  \bibinfo {author} {\bibfnamefont {J.}~\bibnamefont {And{\'e}n}}, \bibinfo
  {author} {\bibfnamefont {J.}~\bibnamefont {Blaschke}},\ and\ \bibinfo
  {author} {\bibfnamefont {A.~H.}\ \bibnamefont {Barnett}},\ }in\ \href
  {https://doi.org/10.1109/IPDPSW52791.2021.00105} {\emph {\bibinfo {booktitle}
  {2021 IEEE International Parallel and Distributed Processing Symposium
  Workshops (IPDPSW)}}}\ (\bibinfo {organization} {IEEE},\ \bibinfo {year}
  {2021})\ pp.\ \bibinfo {pages} {688--697},\ \Eprint
  {https://arxiv.org/abs/2102.08463} {arXiv:2102.08463} \BibitemShut {NoStop}%
\bibitem [{\citenamefont {Springel}\ \emph {et~al.}(2021)\citenamefont
  {Springel}, \citenamefont {Pakmor}, \citenamefont {Zier},\ and\ \citenamefont
  {Reinecke}}]{Gadget4}%
  \BibitemOpen
  \bibfield  {author} {\bibinfo {author} {\bibfnamefont {V.}~\bibnamefont
  {Springel}}, \bibinfo {author} {\bibfnamefont {R.}~\bibnamefont {Pakmor}},
  \bibinfo {author} {\bibfnamefont {O.}~\bibnamefont {Zier}},\ and\ \bibinfo
  {author} {\bibfnamefont {M.}~\bibnamefont {Reinecke}},\ }\href
  {https://doi.org/10.1093/mnras/stab1855} {\bibfield  {journal} {\bibinfo
  {journal} {\mnras}\ }\textbf {\bibinfo {volume} {506}},\ \bibinfo {pages}
  {2871} (\bibinfo {year} {2021})},\ \Eprint {https://arxiv.org/abs/2010.03567}
  {arXiv:2010.03567 [astro-ph.IM]} \BibitemShut {NoStop}%
\bibitem [{\citenamefont {{Brenier}}\ \emph {et~al.}(2003)\citenamefont
  {{Brenier}}, \citenamefont {{Frisch}}, \citenamefont {{H{\'e}non}},
  \citenamefont {{Loeper}}, \citenamefont {{Matarrese}}, \citenamefont
  {{Mohayaee}},\ and\ \citenamefont {{Sobolevski{\u{i}}}}}]{Brenier:2003}%
  \BibitemOpen
  \bibfield  {author} {\bibinfo {author} {\bibfnamefont {Y.}~\bibnamefont
  {{Brenier}}}, \bibinfo {author} {\bibfnamefont {U.}~\bibnamefont {{Frisch}}},
  \bibinfo {author} {\bibfnamefont {M.}~\bibnamefont {{H{\'e}non}}}, \bibinfo
  {author} {\bibfnamefont {G.}~\bibnamefont {{Loeper}}}, \bibinfo {author}
  {\bibfnamefont {S.}~\bibnamefont {{Matarrese}}}, \bibinfo {author}
  {\bibfnamefont {R.}~\bibnamefont {{Mohayaee}}},\ and\ \bibinfo {author}
  {\bibfnamefont {A.}~\bibnamefont {{Sobolevski{\u{i}}}}},\ }\href
  {https://doi.org/10.1046/j.1365-2966.2003.07106.x} {\bibfield  {journal}
  {\bibinfo  {journal} {\mnras}\ }\textbf {\bibinfo {volume} {346}},\ \bibinfo
  {pages} {501} (\bibinfo {year} {2003})},\ \Eprint
  {https://arxiv.org/abs/astro-ph/0304214} {arXiv:astro-ph/0304214 [astro-ph]}
  \BibitemShut {NoStop}%
\bibitem [{\citenamefont {{Fasiello}}\ \emph
  {et~al.}(2022{\natexlab{b}})\citenamefont {{Fasiello}}, \citenamefont
  {{Fujita}},\ and\ \citenamefont {{Vlah}}}]{Fasiello:2022}%
  \BibitemOpen
  \bibfield  {author} {\bibinfo {author} {\bibfnamefont {M.}~\bibnamefont
  {{Fasiello}}}, \bibinfo {author} {\bibfnamefont {T.}~\bibnamefont
  {{Fujita}}},\ and\ \bibinfo {author} {\bibfnamefont {Z.}~\bibnamefont
  {{Vlah}}},\ }\href {https://doi.org/10.1103/PhysRevD.106.123504} {\bibfield
  {journal} {\bibinfo  {journal} {\prd}\ }\textbf {\bibinfo {volume} {106}},\
  \bibinfo {eid} {123504} (\bibinfo {year} {2022}{\natexlab{b}})},\ \Eprint
  {https://arxiv.org/abs/2205.10026} {arXiv:2205.10026 [astro-ph.CO]}
  \BibitemShut {NoStop}%
\bibitem [{\citenamefont {{Nadkarni-Ghosh}}\ and\ \citenamefont
  {{Chernoff}}(2011)}]{Nadkarni-Ghosh:2011}%
  \BibitemOpen
  \bibfield  {author} {\bibinfo {author} {\bibfnamefont {S.}~\bibnamefont
  {{Nadkarni-Ghosh}}}\ and\ \bibinfo {author} {\bibfnamefont {D.~F.}\
  \bibnamefont {{Chernoff}}},\ }\href
  {https://doi.org/10.1111/j.1365-2966.2010.17529.x} {\bibfield  {journal}
  {\bibinfo  {journal} {\mnras}\ }\textbf {\bibinfo {volume} {410}},\ \bibinfo
  {pages} {1454} (\bibinfo {year} {2011})},\ \Eprint
  {https://arxiv.org/abs/1005.1217} {arXiv:1005.1217 [astro-ph.CO]}
  \BibitemShut {NoStop}%
\bibitem [{\citenamefont {{Rampf}}\ and\ \citenamefont
  {{Hahn}}(2023)}]{Rampf:2023}%
  \BibitemOpen
  \bibfield  {author} {\bibinfo {author} {\bibfnamefont {C.}~\bibnamefont
  {{Rampf}}}\ and\ \bibinfo {author} {\bibfnamefont {O.}~\bibnamefont
  {{Hahn}}},\ }\href {https://doi.org/10.1103/PhysRevD.107.023515} {\bibfield
  {journal} {\bibinfo  {journal} {\prd}\ }\textbf {\bibinfo {volume} {107}},\
  \bibinfo {eid} {023515} (\bibinfo {year} {2023})},\ \Eprint
  {https://arxiv.org/abs/2211.02053} {arXiv:2211.02053 [astro-ph.CO]}
  \BibitemShut {NoStop}%
\bibitem [{\citenamefont {Rampf}\ \emph {et~al.}(2023)\citenamefont {Rampf},
  \citenamefont {Saga}, \citenamefont {Taruya},\ and\ \citenamefont
  {Colombi}}]{2023arXiv230312832R}%
  \BibitemOpen
  \bibfield  {author} {\bibinfo {author} {\bibfnamefont {C.}~\bibnamefont
  {Rampf}}, \bibinfo {author} {\bibfnamefont {S.}~\bibnamefont {Saga}},
  \bibinfo {author} {\bibfnamefont {A.}~\bibnamefont {Taruya}},\ and\ \bibinfo
  {author} {\bibfnamefont {S.}~\bibnamefont {Colombi}},\ }\href
  {https://doi.org/10.1103/PhysRevD.108.103513} {\bibfield  {journal} {\bibinfo
   {journal} {Phys. Rev. D}\ }\textbf {\bibinfo {volume} {108}},\ \bibinfo
  {pages} {103513} (\bibinfo {year} {2023})},\ \Eprint
  {https://arxiv.org/abs/2303.12832} {arXiv:2303.12832 [astro-ph.CO]}
  \BibitemShut {NoStop}%
\bibitem [{\citenamefont {Crocce}\ \emph {et~al.}(2006)\citenamefont {Crocce},
  \citenamefont {Pueblas},\ and\ \citenamefont {Scoccimarro}}]{Crocce2006}%
  \BibitemOpen
  \bibfield  {author} {\bibinfo {author} {\bibfnamefont {M.}~\bibnamefont
  {Crocce}}, \bibinfo {author} {\bibfnamefont {S.}~\bibnamefont {Pueblas}},\
  and\ \bibinfo {author} {\bibfnamefont {R.}~\bibnamefont {Scoccimarro}},\
  }\href {https://doi.org/10.1111/j.1365-2966.2006.11040.x} {\bibfield
  {journal} {\bibinfo  {journal} {\mnras}\ }\textbf {\bibinfo {volume} {373}},\
  \bibinfo {pages} {369} (\bibinfo {year} {2006})},\ \Eprint
  {https://arxiv.org/abs/astro-ph/0606505} {arXiv:astro-ph/0606505 [astro-ph]}
  \BibitemShut {NoStop}%
\bibitem [{\citenamefont {{Abel}}\ \emph {et~al.}(2012)\citenamefont {{Abel}},
  \citenamefont {{Hahn}},\ and\ \citenamefont {{Kaehler}}}]{Abel:2012}%
  \BibitemOpen
  \bibfield  {author} {\bibinfo {author} {\bibfnamefont {T.}~\bibnamefont
  {{Abel}}}, \bibinfo {author} {\bibfnamefont {O.}~\bibnamefont {{Hahn}}},\
  and\ \bibinfo {author} {\bibfnamefont {R.}~\bibnamefont {{Kaehler}}},\ }\href
  {https://doi.org/10.1111/j.1365-2966.2012.21754.x} {\bibfield  {journal}
  {\bibinfo  {journal} {\mnras}\ }\textbf {\bibinfo {volume} {427}},\ \bibinfo
  {pages} {61} (\bibinfo {year} {2012})},\ \Eprint
  {https://arxiv.org/abs/1111.3944} {arXiv:1111.3944 [astro-ph.CO]}
  \BibitemShut {NoStop}%
\bibitem [{\citenamefont {{Shandarin}}\ \emph {et~al.}(2012)\citenamefont
  {{Shandarin}}, \citenamefont {{Habib}},\ and\ \citenamefont
  {{Heitmann}}}]{Shandarin:2012}%
  \BibitemOpen
  \bibfield  {author} {\bibinfo {author} {\bibfnamefont {S.}~\bibnamefont
  {{Shandarin}}}, \bibinfo {author} {\bibfnamefont {S.}~\bibnamefont
  {{Habib}}},\ and\ \bibinfo {author} {\bibfnamefont {K.}~\bibnamefont
  {{Heitmann}}},\ }\href {https://doi.org/10.1103/PhysRevD.85.083005}
  {\bibfield  {journal} {\bibinfo  {journal} {\prd}\ }\textbf {\bibinfo
  {volume} {85}},\ \bibinfo {eid} {083005} (\bibinfo {year} {2012})},\ \Eprint
  {https://arxiv.org/abs/1111.2366} {arXiv:1111.2366 [astro-ph.CO]}
  \BibitemShut {NoStop}%
\bibitem [{\citenamefont {{Hahn}}\ \emph {et~al.}(2013)\citenamefont {{Hahn}},
  \citenamefont {{Abel}},\ and\ \citenamefont {{Kaehler}}}]{Hahn:2013}%
  \BibitemOpen
  \bibfield  {author} {\bibinfo {author} {\bibfnamefont {O.}~\bibnamefont
  {{Hahn}}}, \bibinfo {author} {\bibfnamefont {T.}~\bibnamefont {{Abel}}},\
  and\ \bibinfo {author} {\bibfnamefont {R.}~\bibnamefont {{Kaehler}}},\ }\href
  {https://doi.org/10.1093/mnras/stt1061} {\bibfield  {journal} {\bibinfo
  {journal} {\mnras}\ }\textbf {\bibinfo {volume} {434}},\ \bibinfo {pages}
  {1171} (\bibinfo {year} {2013})},\ \Eprint {https://arxiv.org/abs/1210.6652}
  {arXiv:1210.6652 [astro-ph.CO]} \BibitemShut {NoStop}%
\bibitem [{\citenamefont {{Sousbie}}\ and\ \citenamefont
  {{Colombi}}(2016)}]{Sousbie:2016}%
  \BibitemOpen
  \bibfield  {author} {\bibinfo {author} {\bibfnamefont {T.}~\bibnamefont
  {{Sousbie}}}\ and\ \bibinfo {author} {\bibfnamefont {S.}~\bibnamefont
  {{Colombi}}},\ }\href {https://doi.org/10.1016/j.jcp.2016.05.048} {\bibfield
  {journal} {\bibinfo  {journal} {\jcp}\ }\textbf {\bibinfo {volume} {321}},\
  \bibinfo {pages} {644} (\bibinfo {year} {2016})},\ \Eprint
  {https://arxiv.org/abs/1509.07720} {arXiv:1509.07720 [physics.comp-ph]}
  \BibitemShut {NoStop}%
\bibitem [{\citenamefont {Wang}\ and\ \citenamefont
  {White}(2007)}]{10.1111/j.1365-2966.2007.12053.x}%
  \BibitemOpen
  \bibfield  {author} {\bibinfo {author} {\bibfnamefont {J.}~\bibnamefont
  {Wang}}\ and\ \bibinfo {author} {\bibfnamefont {S.~D.~M.}\ \bibnamefont
  {White}},\ }\href {https://doi.org/10.1111/j.1365-2966.2007.12053.x}
  {\bibfield  {journal} {\bibinfo  {journal} {\mnras}\ }\textbf {\bibinfo
  {volume} {380}},\ \bibinfo {pages} {93} (\bibinfo {year} {2007})},\ \Eprint
  {https://arxiv.org/abs/astro-ph/0702575} {arXiv:astro-ph/0702575}
  \BibitemShut {NoStop}%
\bibitem [{\citenamefont {Romeo}\ \emph {et~al.}(2008)\citenamefont {Romeo},
  \citenamefont {Agertz}, \citenamefont {Moore},\ and\ \citenamefont
  {Stadel}}]{Romeo_2008}%
  \BibitemOpen
  \bibfield  {author} {\bibinfo {author} {\bibfnamefont {A.~B.}\ \bibnamefont
  {Romeo}}, \bibinfo {author} {\bibfnamefont {O.}~\bibnamefont {Agertz}},
  \bibinfo {author} {\bibfnamefont {B.}~\bibnamefont {Moore}},\ and\ \bibinfo
  {author} {\bibfnamefont {J.}~\bibnamefont {Stadel}},\ }\href
  {https://doi.org/10.1086/591236} {\bibfield  {journal} {\bibinfo  {journal}
  {\apj}\ }\textbf {\bibinfo {volume} {686}},\ \bibinfo {pages} {1} (\bibinfo
  {year} {2008})}\BibitemShut {NoStop}%
\bibitem [{\citenamefont {Springel}(2005)}]{Gadget2}%
  \BibitemOpen
  \bibfield  {author} {\bibinfo {author} {\bibfnamefont {V.}~\bibnamefont
  {Springel}},\ }\href {https://doi.org/10.1111/j.1365-2966.2005.09655.x}
  {\bibfield  {journal} {\bibinfo  {journal} {\mnras}\ }\textbf {\bibinfo
  {volume} {364}},\ \bibinfo {pages} {1105} (\bibinfo {year} {2005})},\ \Eprint
  {https://arxiv.org/abs/0505010} {arXiv:0505010 [astro-ph]} \BibitemShut
  {NoStop}%
\bibitem [{\citenamefont {Hahn}\ and\ \citenamefont {Abel}(2011)}]{Hahn2011}%
  \BibitemOpen
  \bibfield  {author} {\bibinfo {author} {\bibfnamefont {O.}~\bibnamefont
  {Hahn}}\ and\ \bibinfo {author} {\bibfnamefont {T.}~\bibnamefont {Abel}},\
  }\href {https://doi.org/10.1111/j.1365-2966.2011.18820.x} {\bibfield
  {journal} {\bibinfo  {journal} {\mnras}\ }\textbf {\bibinfo {volume} {415}},\
  \bibinfo {pages} {2101} (\bibinfo {year} {2011})},\ \Eprint
  {https://arxiv.org/abs/1103.6031} {arXiv:1103.6031 [astro-ph.CO]}
  \BibitemShut {NoStop}%
\bibitem [{\citenamefont {Sefusatti}\ \emph {et~al.}(2016)\citenamefont
  {Sefusatti}, \citenamefont {Crocce}, \citenamefont {Scoccimarro},\ and\
  \citenamefont {Couchman}}]{sefusatti2016accurate}%
  \BibitemOpen
  \bibfield  {author} {\bibinfo {author} {\bibfnamefont {E.}~\bibnamefont
  {Sefusatti}}, \bibinfo {author} {\bibfnamefont {M.}~\bibnamefont {Crocce}},
  \bibinfo {author} {\bibfnamefont {R.}~\bibnamefont {Scoccimarro}},\ and\
  \bibinfo {author} {\bibfnamefont {H.~M.}\ \bibnamefont {Couchman}},\ }\href
  {https://doi.org/10.1093/mnras/stw1229} {\bibfield  {journal} {\bibinfo
  {journal} {\mnras}\ }\textbf {\bibinfo {volume} {460}},\ \bibinfo {pages}
  {3624} (\bibinfo {year} {2016})},\ \Eprint {https://arxiv.org/abs/1512.07295}
  {arXiv:1512.07295 [astro-ph.CO]} \BibitemShut {NoStop}%
\bibitem [{\citenamefont {{Chen}}\ \emph {et~al.}(1974)\citenamefont {{Chen}},
  \citenamefont {{Bruce Langdon}},\ and\ \citenamefont
  {{Birdsall}}}]{Chen:1974}%
  \BibitemOpen
  \bibfield  {author} {\bibinfo {author} {\bibfnamefont {L.}~\bibnamefont
  {{Chen}}}, \bibinfo {author} {\bibfnamefont {A.}~\bibnamefont {{Bruce
  Langdon}}},\ and\ \bibinfo {author} {\bibfnamefont {C.~K.}\ \bibnamefont
  {{Birdsall}}},\ }\href {https://doi.org/10.1016/0021-9991(74)90014-X}
  {\bibfield  {journal} {\bibinfo  {journal} {\jcp}\ }\textbf {\bibinfo
  {volume} {14}},\ \bibinfo {pages} {200} (\bibinfo {year} {1974})}\BibitemShut
  {NoStop}%
\bibitem [{\citenamefont {Appel}(1985)}]{appel1985efficient}%
  \BibitemOpen
  \bibfield  {author} {\bibinfo {author} {\bibfnamefont {A.~W.}\ \bibnamefont
  {Appel}},\ }\href@noop {} {\bibfield  {journal} {\bibinfo  {journal} {SIAM J.
  Sci. Comput.}\ }\textbf {\bibinfo {volume} {6}},\ \bibinfo {pages} {85}
  (\bibinfo {year} {1985})}\BibitemShut {NoStop}%
\bibitem [{\citenamefont {Barnes}\ and\ \citenamefont
  {Hut}(1986)}]{barnes1986hierarchical}%
  \BibitemOpen
  \bibfield  {author} {\bibinfo {author} {\bibfnamefont {J.}~\bibnamefont
  {Barnes}}\ and\ \bibinfo {author} {\bibfnamefont {P.}~\bibnamefont {Hut}},\
  }\href@noop {} {\bibfield  {journal} {\bibinfo  {journal} {Nature}\ }\textbf
  {\bibinfo {volume} {324}},\ \bibinfo {pages} {446} (\bibinfo {year}
  {1986})}\BibitemShut {NoStop}%
\bibitem [{\citenamefont {Uhlemann}\ \emph {et~al.}(2019)\citenamefont
  {Uhlemann}, \citenamefont {Rampf}, \citenamefont {Gosenca},\ and\
  \citenamefont {Hahn}}]{Uhlemann:2018gzz}%
  \BibitemOpen
  \bibfield  {author} {\bibinfo {author} {\bibfnamefont {C.}~\bibnamefont
  {Uhlemann}}, \bibinfo {author} {\bibfnamefont {C.}~\bibnamefont {Rampf}},
  \bibinfo {author} {\bibfnamefont {M.}~\bibnamefont {Gosenca}},\ and\ \bibinfo
  {author} {\bibfnamefont {O.}~\bibnamefont {Hahn}},\ }\href
  {https://doi.org/10.1103/PhysRevD.99.083524} {\bibfield  {journal} {\bibinfo
  {journal} {Phys. Rev. D}\ }\textbf {\bibinfo {volume} {99}},\ \bibinfo
  {pages} {083524} (\bibinfo {year} {2019})},\ \Eprint
  {https://arxiv.org/abs/1812.05633} {arXiv:1812.05633 [astro-ph.CO]}
  \BibitemShut {NoStop}%
\end{thebibliography}
\end{document}